\documentclass{emulateapj}
\usepackage{graphicx}
\usepackage{mathtools}

\newcommand \mic {$\mu$m}
\newcommand \lsun {L$_{\odot}$}
\newcommand \msun {M$_{\odot}$}
\shortauthors{Staguhn \etal} \shorttitle{The GISMO Deep Field (GDF)} \tighten

\def\etal   {et~al.~}

\slugcomment{Rev. 06-07-14}

\begin{document}
\title{The GISMO 2-millimeter Deep Field in GOODS-N}

\author{Johannes G. Staguhn\altaffilmark{1,2}, Attila Kov{\'a}cs\altaffilmark{3,4},
Richard G. Arendt\altaffilmark{2,5}, Dominic J. Benford\altaffilmark{2}, Roberto Decarli\altaffilmark{6}, Eli Dwek\altaffilmark{2}, Dale J. Fixsen\altaffilmark{2,7}, Gene C. Hilton\altaffilmark{8}, Kent D. Irwin\altaffilmark{8,9}, Christine A. Jhabvala\altaffilmark{2}, Alexander Karim\altaffilmark{10,11} Samuel Leclercq\altaffilmark{12}, Stephen F. Maher\altaffilmark{2,13},  Timothy M. Miller\altaffilmark{2}, S.Harvey Moseley\altaffilmark{2}, Elmer H. Sharp\altaffilmark{2,14}, Fabian Walter\altaffilmark{6}, Edward J. Wollack\altaffilmark{2}}

\altaffiltext{1}{The Henry A. Rowland Department of Physics and Astronomy, Johns Hopkins University, 3400 N. Charles Street, Baltimore, MD 21218, USA}
\altaffiltext{2}{Observational Cosmology Lab, Code 665, NASA Goddard Space Flight Center, Greenbelt, MD 20771, USA}
\altaffiltext{3}{California Institute of Technology 301-17, 1200 E California Blvd, Pasadena, CA 91125} 
\altaffiltext{4}{Institute for Astrophysics, University of Minnesota, 116 Church St SE, Minneapolis, MN 55455, USA}
\altaffiltext{5}{CRESST, University of Maryland -- Baltimore County, Baltimore, MD 21250, USA}
\altaffiltext{6}{Max-Planck-Institute f\"ur Astronomie, K\"onigstuhl 17, D-69117, Heidelberg, Germany}
\altaffiltext{7}{CRESST, University of Maryland, College Park, College Park, MD 20742, USA}
\altaffiltext{8}{NIST Quantum Devices Group, 325 Broadway Mailcode 817.03, Boulder, CO 80305, USA}
\altaffiltext{9}{Dept. of Physics, Stanford University, Stanford, CA 94305, USA}
\altaffiltext{10}{Department of Physics, Durham University, South Road, Durham DH1 3LE, UK}
\altaffiltext{11}{Argelander-Institut fŸr Astronomie, University of Bonn, Auf dem H\"ugel 71, 53121 Bonn, Germany}
\altaffiltext{12}{Institut de Radio Astronomie Millim\'etrique, 300 Rue de la Piscine, 38406 Saint Martin d`Heres, France}
\altaffiltext{13}{Science Systems and Applications, Inc., 10210 Greenbelt Rd, Suite 600, Lanham, MD 20706, USA}
\altaffiltext{14}{Global Science \& Technology, Inc., 7855 Walker Drive, Suite 200, Greenbelt, MD 20770, USA}

\begin{abstract}
We present deep continuum observations using the GISMO camera at a wavelength of 2~mm centered on the
Hubble Deep Field (HDF) in the GOODS-N field. These are the first deep field
observations ever obtained at this wavelength. The $1\sigma$ sensitivity in the
innermost $\sim4'$ of the $7'$ diameter map is $\sim135$ $\mu$Jy/beam, a factor
of three higher in flux/beam sensitivity than the deepest available SCUBA
850 $\micron$ observations, and almost a factor of four higher in flux / beam sensitivity than the combined
MAMBO/AzTEC 1.2 mm observations of this region. Our source extraction algorithm
identifies 12 sources directly, and another 3 through correlation with known
sources at 1.2~mm and 850 $\micron$. 
Five of the directly detected GISMO sources
have counterparts in the MAMBO/AzTEC catalog, and four of those also
have SCUBA counterparts. 
HDF850.1, one of the first blank-field detected submillimeter galaxies,
is now detected at 2~mm.
The median redshift of all sources
with counterparts of known redshifts is $\tilde{z} = 2.91\pm0.94$.
Statistically, the detections are most likely real for 5 of the
seven 2 mm 
sources without shorter wavelength counterparts, while the probability for none of them being real is negligible. 


\end{abstract}

\keywords{galaxies: evolution -- galaxies: high-redshift -- galaxies: luminosity function -- galaxies: photometry -- galaxies: starburst -- infrared: galaxies}

\section{Introduction}
Submillimeter and millimeter observations have revealed the existence of a population of previously unknown high-redshift dust-enshrouded starburst galaxies. Virtually all their  stellar UV and optical radiation is absorbed and reradiated by the dust at infrared (IR) wavelengths. They are among the most luminous galaxies in the universe, and their relative contribution to the galaxy number counts and co-moving luminosity density increases with redshift (e.g. \citet{2012ApJ...747L..31S}). 

 The most massive galaxies are predicted to be at the center of galaxy clusters that reside in the most massive dark matter halos. Surveys that map their distribution with redshift will therefore reveal the epochs of cluster formation in the early universe. For example, follow up observations of two submillimeter galaxies at optical and near-IR wavelengths have shown that they are members of protoclusters that formed at $z\approx 5$ (AzTEC-3: \citet{2011Natur.470..233C}, HDF850.1: \citet{2012Natur.486..233W}). A survey of dusty starbursts is also essential for determining the obscured cosmic star formation rate at high redshift, and for understanding the formation and evolution of dust in these objects.  

The advantage of using  (sub)millimeter wavelength observations to search for these objects stems from the fact that starburst galaxies have typical dust temperatures of 35~K, so that their IR spectrum peaks  at $\sim 90$~\mic. Submm--mm observations therefore trace the Rayleigh-Jeans part of their spectrum, and benefit from the fact that the decrease in flux from high redshift objects is largely offset by the negative K-correction.  

\begin{figure}
\includegraphics[angle=0,width=3.3in]{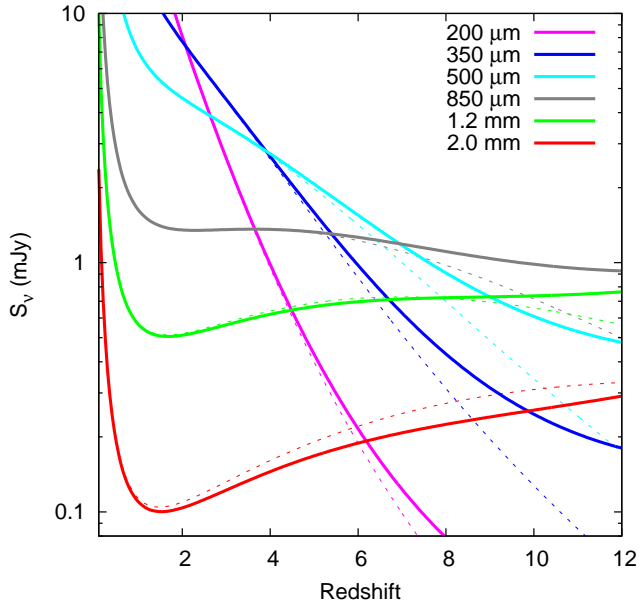}
\caption{Redshift-dependent flux density, measured against the CMB, of a galaxy with 
fixed dust mass $M_{\mbox{d}} = 10^8 M_\odot$,  dust emissivity 
index of $\beta$=1.5, i.e. with a FIR luminosity of $L \sim10^{12} L_\odot$, shown for a variety of wavelengths, based on the multi-temperature empirical dust models of Kov\'{a}cs et al.\ 2010). To radiate the same luminosity against an increasingly warmer CMB in earlier epochs, the cold dust temperature ($T_0$=35\,K) must rise as $T_d^{4+\beta_{\rm eff}} = T_{\rm CMB}^{4 + \beta_{\rm eff}} + T_0^{4 + \beta_{\rm eff}}$ (the effective dust emissivity, $\beta_{\rm eff}$, is defined in Kov\'{a}cs et al.\ 2010). For comparison, the dashed lines show the same if the CMB heating is ignored. Note, that the 
observed flux density at 2~mm wavelengths increases monotonically and steeply as a function of 
redshift for $z>$1. \\
\label{K-correct}}
\end{figure}

Figure~\ref{K-correct} depicts the flux of a typical dusty submillimeter galaxy versus redshift (solid lines). This starburst galaxy is characterized by an IR luminosity of $10^{12}$~\lsun\  and a dust mass of $10^8$~\msun. The dust was assumed to have a $\kappa(\lambda) \propto \lambda^{-\beta}$ mass absorption coefficient with a spectral index of $\beta = 1.5$, and a temperature of 35~K. 
An interesting effect at high redshifts is the fact that dust heating by the cosmic microwave background (CMB) becomes comparable to the heating by ambient starlight. When accounting for both sources of heating, the actual dust temperature can be expressed as $T_d = (T_0^{4+\beta} + T_{CMB}^{4+\beta})^{1/(4+\beta)}$, where $T_{CMB}=2.73(1+z)$ is the CMB temperature at redshift $z$, and $T_0$ is the dust temperature when heated by starlight alone. GISMO (the the Goddard IRAM 2 Millimeter Observer) observations are a differential measurement of the flux from a galaxy against the CMB. The observed galaxy spectrum in such measurement is thus given by:
\begin{equation}
\label{ F_nu}
F_{\nu}(\lambda) = 4\pi M_d\ \kappa(\lambda)\, \left[B_{\nu}(\lambda, T_d) - B_{\nu}(\lambda, T_{CMB}\right]
\end{equation}   
which cannot be characterized by a single blackbody with a simple $\lambda^{-\beta}$ emissivity law.
The dotted lines in the figure show the fluxes that would be measured if the CMB radiation were not present.

The figure shows that the 2~mm fluxes tend to be lower than those at the shorter wavelengths. However, the rising 2~mm flux with redshift provides the strongest bias towards the detection of high redshift galaxies. Furthermore, the atmospheric transmission is higher, and the atmospheric background noise is lower at 2~mm than at shorter wavelengths.  

We have developed the GISMO instrument that utilizes a near background-limited detector to fully exploit the advantages of the 2 millimeter window. Here we report the first deep survey conducted with GISMO centered on the Hubble Deep Field North (HDF-N).  The Hubble Deep Field North (HDF-N) is one of the best studied regions in the sky. Its sky coverage is one HST WFPC2 pointing, i.e. $\sim 2.5' \times 2.5'$, which is less than the $2' \times 4'$ instantaneous field of view of the GISMO array. The HDF is located in the greater GOODS-N region, which has also been studied in exquisite detail over the last decade at many wavelengths, X-rays: \citet{Bra08}, UV: \citet{2006AJ....132..853T}, Optical: \citet{2004ApJ...600L..93G}, optical spectroscopy: \citet{2004AJ....127.3137C}, Near-Infrared: \citet{2006NewAR..50..127Y}, Mid-Infrared: \citet{2006MNRAS.371.1891R}, Far-Infrared:\cite{2003MNRAS.344..385B}, \citet{2006ApJ...647L...9F}, \citet{2008MNRAS.391.1227P}, Radio: \citet{2010ApJS..188..178M}. In the (sub-)millimeter regime, the HDF and GOODS-N have been studied by the (sub-)millimeter cameras SCUBA,  MAMBO, and AzTEC in the past (\citet{1998Natur.394..241H}, \citet{2005MNRAS.358..149P}; \citet{2008MNRAS.389.1489G}, \citet{2011MNRAS.410.2749P}. The currently available data of the full HDF at 1~mm reach $1-\sigma$ sensitivities of 0.5 mJy/beam (MAMBO observations combined with AzTEC observations, \citep{2011MNRAS.410.2749P}, the SCUBA ``super''--map \citep{2005MNRAS.358..149P} reaches a peak depth of 0.4 mJy\,beam$^{-1}$ at $850~\mu$m, however the sensitivity varies significantly over the observed area in the field. 

The paper is organized as follows: We first describe the instrument and its characteristics in \S2. In \S3 we describe  observations and the data reduction. In \S4 we describe the source extraction and its results, and present simulations used to characterize the
data and to evaluate the completeness and reliability of the extracted sources. \S5 presents the 2~mm number counts and the analysis of the properties of select individual sources.

\section{The GISMO 2~mm Camera}

Continuum observations in the 2\,mm atmospheric window have not been astronomically explored from the ground to the same degree as has been done at shorter wavelengths (1~mm or less), 
except for Sunyaev-Zel'dovich (S-Z) observations with dedicated  6 to 10~m class telescopes (\citet{2011ApJS..194...41S}, \citet{2011PASP..123..568C}, \citet{2006NewAR..50..960D}). 
The reason for this is predominantly of technical nature, in particular the very demanding requirements on the noise performance of a background-limited camera operating in this low opacity atmospheric window. In order to provide background-limited observations in the 2~mm window at a good mountain site such as the the IRAM 30~m telescope on Pico Veleta, the required sensitivity, expressed in Noise Equivalent Power (NEP), for the detectors is $\sim 5\times10^{-17}$ W $\sqrt{\mbox{s}}$ \citep{2006SPIE.6275E..44S}, a requirement that is met by our ``high'' temperature ($T_c$ = 450 mK) Transition Edge Sensor (TES) detectors.  Consequently we have  proposed and built a 2~millimeter wavelength bolometer camera, the Goddard-IRAM Superconducting 2 Millimeter Observer (GISMO, \citet{2008SPIE.7020E...3S}) for astronomical observations at the IRAM 30\,m telescope on Pico Veleta, Spain \citep{1987A&A...175..319B}. GISMO uses a compact optical design \citep{2008SPIE.7020E..66S} and uses an $8\times16$ array of close-packed, high sensitivity TES bolometers with a pixel size of $2\times2$ mm$^2$ \citep{2008SPIE.7020E..56B}, which was built in the Detector Development Laboratory at NASA/GSFC \citep{2008JLTP..151..266A}. The array architecture is based on the Backshort Under Grid  (BUG) design \citep{2006SPIE.6275E...9A}. GISMO's bandpass is centered on 150 GHz and has a fractional bandwidth of 20\%. The superconducting bolometers are read out by SQUID time domain multiplexers from NIST/Boulder \citep{2002AIPC..605..301I}. 
This design is scalable to kilopixel size arrays for future ground-based, suborbital and space-based X-ray and far-infrared through millimeter cameras (e.g. \citet{2012SPIE.8452E..0TS}).

\section{Observations \& Reduction}

The GISMO Deep Field (GDF) observations of the HDF-N were obtained between April 13 and 18, 2011, and on April 11, 12 and 23, 2012. The total integration time was $t\sim39$ hours, however 2/3 of those observations were obtained with GISMO's lower sensitivity during the Spring 2011 run (see section \ref{sec:inst_perf}). The FWHM of GISMO's beam is $17.5"$.

\subsection{Data Reduction}

The data were reduced, using CRUSH\footnote{http://www.submm.caltech.edu/$\sim$sharc/crush} \citep{2008SPIE.7020E..45K}, which is the standard reduction software for the GISMO camera. CRUSH is open-source and available in the public domain. The data reduction tool of CRUSH consists of a highly configurable pipeline, which uses a series of statistical estimators in an iterated scheme to separate the astronomical signals from the bright and variable atmospheric background and various correlated instrumental noise signals. It determines proper noise weights for each sample in the time series, removes glitches, identifies bad pixels and other unusable data, and determines the relevant relative gains. It also applies appropriate filters for $1/f$-type noise, and other non-white detector noise profiles. For the detection of point sources, the resulting "deep"-mode maps are spatially filtered above $50''$ FWHM to remove spatially-variant atmospheric residuals. The fluxes in each 10-minute scan are corrected for the line-of-sight atmospheric opacities, based on the IRAM radiometer measurements. Point-source fluxes are also corrected scan-wise for the flux-filtering effect of each and every pipeline step, and for the large-scale structure filtering of the final map. As a result, comparison to point-like calibrator sources (e.g.\ planets and quasars) is straightforward, even if different reduction options are used for these and the science targets. 

\subsection{Calibration}
\begin{figure}
\includegraphics[angle=0,width=3.3in]{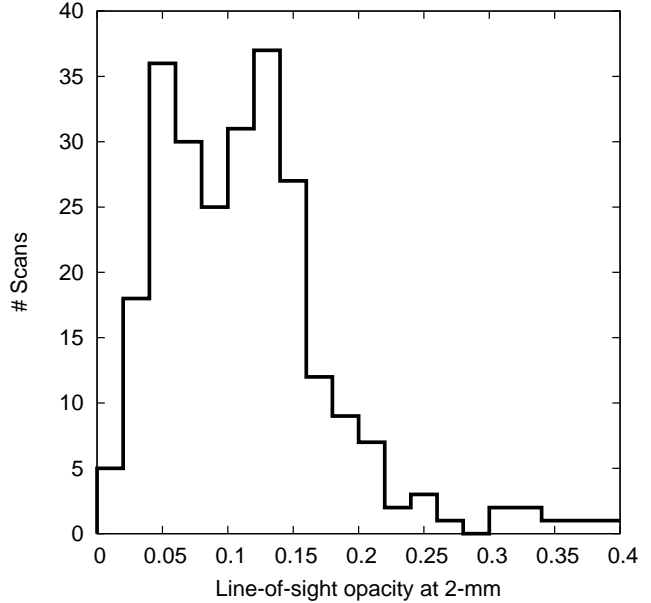}
\caption{Histogram of the 2-mm line of sight opacities for the GDF observations}
\label{opacities}
\end{figure}

Mars, Uranus, and Neptune were observed for primary flux calibration. Of those, measurements of Mars cover the widest range of weather conditions. Using the atmospheric transmission model of the Caltech Submillimeter Observatory\footnote{http://www.submm.caltech.edu/cso/weather/atplot.shtml} and the IRAM 30~m Telescope 225~GHz radiometer readings, we obtain excellent calibration in effectively all weather conditions: a 7\% rms blind calibration up to $\tau_{\rm 225GHz} \sim 1$ is obtained. Note that any model uncertainties due to the different elevation of Mauna Kea, the site of the CSO, and Pico Veleta, will be very small and therefore irrelevant for the accuracy of derived calibration factors. Figure \ref{opacities} shows the histogram of the 2~mm line-of-sight opacities for all data.

\subsection{Pointing \& Astrometric Accuracy}

During the GISMO  observing runs in 2011 and 2012, we obtained a large number of pointing measurements over the entire sky, from which we derived appropriate pointing models according to \citet{1996A&AS..115..379G}. Our pointing models yield $<$3'' rms accuracy in both Az and El directions on all pointing measurements obtained during the two observing runs (424 and 392 individual pointing observations for the 2011 and 2012 observing runs, respectively). Additionally, we frequently checked pointing on nearby quasars during GDF observations. Triggered by a reduction flag, CRUSH will automatically incorporate the measured differential offsets with respect to the pointing model, to further improve pointing accuracy, and to remove most systematic pointing errors in the pointing model, or due to structural deformations of the telescope. The resulting residual pointing errors are expected to be independent and random between independent pointing sessions. Thus, a representative lower bound to the final astrometric accuracy is given by the instantaneous pointing rms ($<$4.2'') divided by the square root of the number of independent pointing sessions spanning the observations. In our case, approximately 30 independent pointing sessions bracket the GDF observations. Therefore, the astrometric accuracy of our map (notwithstanding the inherent positional uncertainties of any detections) could be as low as 0.8'' rms, or somewhat higher in the presence of systematics errors, which are not eliminated by the use of nearby pointing measurements.




\subsection{Instrument Performance}
\label{sec:inst_perf}

The noise equivalent flux density (NEFD) of measurements during the 2011 run was typically 15-17 mJy$\sqrt{\mbox{s}}$, under most weather conditions. The obtained sensitivity at that time was mainly limited by a neutral density filter with 40\% transmission. This filter was needed, since there was a significant amount of  THz light  scattered into the GISMO beam by the low pass filters, which were positioned very close to the entrance window of the dewar. In early 2012 we mounted a 77\,K baffle dewar in front of the  GISMO optical entrance window, which reduces the stray light significantly and eliminates the need for a neutral density filter in the instrument \citep{2012SPIE.8452E..3IS}. As a result of this, the NEFD obtained during the 2012 observing run was typically 10~mJy$\sqrt{\mbox{s}}$.

\subsection{Noise properties of the beam-smoothed map}
\label{sec:smoothnoise}

\begin{figure}
\includegraphics[]{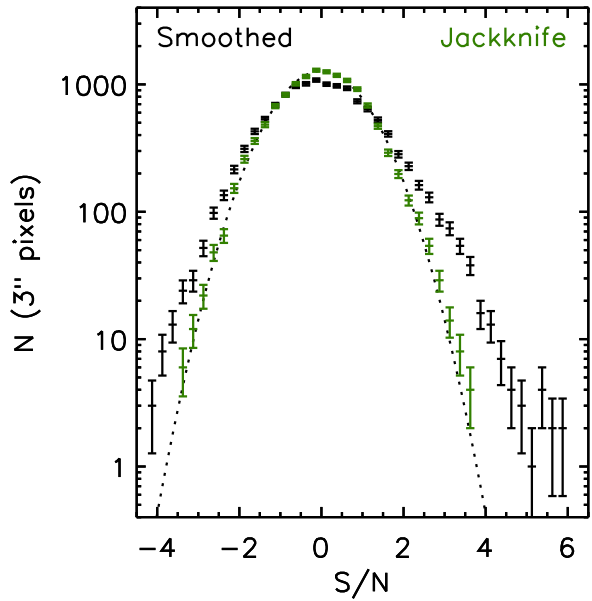}
\caption{Signal to noise histogram for the smoothed and filtered GDF 
data (black), which was used for the source extraction, and shown in Fig.~\ref{GDF}, and for a corresponding time-based random jackknife realization (green). The dotted line shows the expected Gaussian noise distribution, based on the radiometric down-integration of the detector timestream noise, with $\sigma = 1.00$. The close agreement of the jackknifed noise distribution and the Gaussian expectation indicates that our measurement noise is both closely Gaussian in nature, and radiometric. At the same time, the histogram of the regular (non-jackknifed) map exhibits distinct deviations from the Gaussian noise. A symmetric widening is caused by additional noise from unresolved sources (i.e.\ confusion noise) on top of the radiometric measurement noise, while resolved emission sources cause the asymmetric excess on the positive half of the distribution. \label{GDF-hist}} 
\end{figure}

\begin{figure}
\includegraphics[]{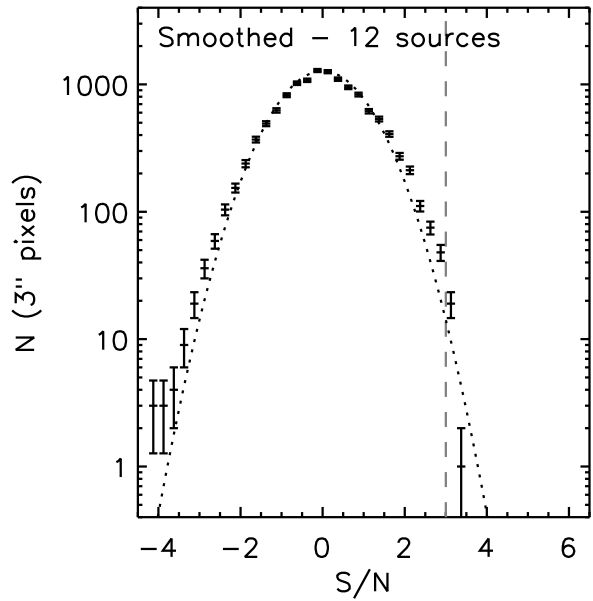}
\caption{Signal to noise histogram of the smoothed and filtered GDF data (Fig. \ref{GDF}), after the 12 blindly identified sources are removed and flagged. The vertical dashed line marks the source extraction threshold, resulting in an abrupt truncation of the histogram above the detection threshold. The dotted line shows the expected Gaussian noise distribution from the jackknifed map with $\sigma = 1.00$. The post-extraction residuals are more closely Gaussian, as expected, albeit still hinting at the presence of further (fainter) resolved and unresolved sources in the field, which manifest as an asymmetric and a symmetric excess, respectively.
\label{GDF-hist-12s}}
\end{figure}

To estimate the noise we randomly multiply each of the individual 10 minute scans by +1 or -1, a method known as ``jackknifing''. This eliminates any stationary noise (including sources, and foreground) but retains random noise, including that from the atmosphere. The histogram of the S/N for the jackknifed, beam-smoothed, and filtered GISMO map is shown in 
Figure~\ref{GDF-hist}. The distribution is well fit by a Gaussian with $\sigma = 1.00$.  The S/N histogram for the 
regular (not jackknifed) smoothed and filtered map (also Fig. \ref{GDF-hist}) shows distinct 
excess deviations from the Gaussian distribution on both extremes of the distribution. When we subtract the 12 detected sources (see section \ref{sec:results}) from the image, 
the S/N histogram does approach the expected noise distribution as shown in Figure \ref{GDF-hist-12s}.
The subtraction of sources (both real and false) causes this histogram to be truncated at the $2.99\sigma$ 
detection limit. Below this limit both the positive and negative sides of the histogram are more closely Gaussian because the effective smoothed and filtered PSF used for the subtraction has both positive
and negative features (main beam and surrounding filter bowl). There is, however, a slight ($\sim 5\%$) symmetric excess remaining in the post-extraction residuals, relative to the jackknifed map. This excess would be consistent with the presence of confusion noise in our map. We cannot, however, quantify the corresponding level of 2-mm confusion noise with any precision because the the noise estimation errors are relatively large given the low number of beams in the field.

\section{Source Extraction \& Simulations}
\subsection{Extraction -- Method and reliability}
\label{sec:stats}

The Gaussian nature of noise in our map (section \ref{sec:smoothnoise}) is a notable feature of the GISMO data and allows us to provide the strongest possible statistical characterization of our source candidates relying exclusively on formal Gaussian statistics. Because we do not detect any deviation from Gaussian noise, nor see any signs of non-radiometric down-integration in the jackknifed maps, we need not worry about potential troublesome statistical biases that could otherwise result from non-Gaussian, or correlated, noise features. 

We extract sources from the beam-smoothed filtered map, which were produced by CRUSH. Beam-smoothing is mathematically equivalent to maximum-likelihood PSF amplitude fitting at every map position \citep{Kovacs-thesis}. I.e., the PSF-smoothed map value and its noise directly provide the amplitude and uncertainty of a fitted PSF at each map position. The effective filtered PSF for GISMO maps reduced in 'deep' mode in CRUSH is accurately modeled as a combination of a 17.5" FWHM Gaussian main beam combined with a negative 50" FWHM Gaussian bowl such that the combined PSF yield zero integral (signifying that we have no DC sensitivity due to the sky-noise removal, and other filtering during the reduction). The 50" FWHM effective PSF bowl is the direct result of explicit large-scale-structure (LSS) filtering during the reduction with a 50" FWHM Gaussian profile. We checked the Gaussian main beam assumption on quasars (detected at high signal-to-noise $>$1000), and confirmed that $>$98\% of the integrated flux inside a $R$=50" circular aperture is recovered in the 17.5" FWHM beam-smoothed peak during night-time observations (i.e. for all of our data). We also checked that the filtered PSF accurately recovers the quasar fluxes, when these were LSS filtered the same as our deep field map, and found no further degradation of photometric accuracy associated with the LSS filtering. Therefore, we are very confident that we have sufficient understanding of the effective PSF in our map, and that the systematic errors of the extracted points source fluxes are kept below 2\%.

The source extraction code we used is part of the CRUSH software package, and is the same source extraction tool that was used and described in \citet{2009ApJ...707.1201W}. It implements an iterated false-detection rate algorithm. Apart from peak position and flux, the algorithm calculates an estimated confidence and an expected cumulative false detection rate for each extracted source. We caution that the confidence levels and false-detection rates are guiding values only, which represent our best statistical estimate without prior knowledge of the 2-mm source counts. A more accurate characterization of confidence levels and/or cumulative false-detection rates would require accurate prior information of the true counts of the 2-mm source population.

\subsubsection{Overview of the CRUSH source extraction tool.}

Here, we offer a concise summary of the approach implemented by CRUSH 'detect' tool, which we used for the source extraction.

The expected false detection rate, i.e. the expected
number of pure noise peaks mistakenly identified as a source, 
is given by $N_f(\Sigma) = N Q(\Sigma)$, where $N$ is the number of 
independent Gaussian variables in the map, and $Q(\Sigma) = 1-P(\Sigma)$
is the complement cumulative Gaussian probability, i.e. the probability 
of measuring a deviation larger than a chosen significance, $\Sigma$.
A smoothed and filtered map with extraction area $A$ contains

\begin{equation}
N \approx  \frac {4.5 A} {1.13~(\Delta_{\rm
smooth}^2 ) } \left(1 -
\frac{\Delta_{\rm smooth}^2} {\Delta_{\rm filter}^2} \right)
\label{}
\end{equation}

independent variables in terms of the FWHM widths of the
Gaussian smoothing ($\Delta_{\rm smooth}$=$\Delta_{\rm beam}$) and the
applied large-scale filtering of the map ($\Delta_{\rm filter}$=50''). The right-hand-side term in the formula accounts for the lost degrees of
freedom due to the explicit spatial filtering of our map. The approximately 4.5
parameters per smoothing beam were determined empirically based on the occurrence of significant noise peaks in simulated noise maps. The formula was verified to yield close to the expected number of false detections in simulated noise maps with varying areas and filtering properties, and with $N_f$ targeted between 0 to 1000. Thus, the above expression will accurately predict the actual false detection rate, as a function of detection threshold, as long as the map noise is known precisely. For our map, with $\sim$321 GISMO beams, a 2.99\,$\sigma$ cut yields $N_f(2.99) \approx 2 $ expected false detections.

Due to the presence of many resolved but undetected sources in the map (asymmetric confusion), our noise estimates are bound to be slightly overestimated (even with the median-noise based estimate used). To our best knowledge, all statistical estimates of map noise, which are based on the observed map itself, will result in overstated noise estimates in the presence of asymmetric confusion (resolved sources below detection). Neither the jackknifed noise, nor radiometric noise, can help offer better estimates, as the extraction noise should include the effect of symmetric confusion (unresolved faint sources) beyond what these can offer. As a result of an inevitably biased noise estimation process, the corresponding false detection rate estimates are slightly above actual, and represent a useful conservative upper bound. This is confirmed by the simulations, presented in section \ref{sec:sims}, which found that if the 2~mm source counts were those of, e.g., \citet{2011A&A...529A...4B} or \citet{2011ApJ...742...24L}, then the actual false detection rate would be 1.34 or 0.55, respectively, vs.\ the expected 2.  However, as we stated earlier, we cannot unbias our noise estimates, or quantify the true false detection rate, without prior knowledge of the true 2-mm source counts, which are not well-constrained at present. Instead, our estimates offer strong upper bounds for the unknown actual false-detection rates.

Each source identified above the significance cut is removed from the map with the smoothed and filtered point-spread function before the extraction proceeds. Subtraction with the filtered PSF allows the detection of further nearby peaks, which may have been previously suppressed by the negative filter bowl surrounding the previous detections. The circular area ($r^2$=$\Delta_{\rm beam}^2 + \Delta_{\rm smooth}^2$) containing the main beam of the detected source is flagged after the extraction, since it no longer contains meaningful information after the removal of the source from within. To ensure that our
catalog is based on the most accurate measure of the map noise and
zero levels, CRUSH estimates the zero level using the mode of the map flux
distribution, and estimates the noise from the median observed
deviation median$(x^2) \approx 0.454937 \sigma^2$). Both measures are relatively robust and reasonably unbiased by the presence of
relatively bright sources, or localized features, in the map.

For each source candidate, CRUSH estimates a detection confidence based on the expected false detection rate $N_f$. According to Poisson statistics, the detection confidence $C$ of a single peak is the
probability that no such peak occurs randomly, i.e. $P_0(N_f) = e^{-N_f}$. This is then further refined to include information from other sources already detected in the map. Thus, if $n$ true sources with apparent significance above
$\Sigma$ are known {\em a priori} to exist in the map, than any
given peak at significance $\Sigma$ may be one of $n$ sources, or one of
the $N_f$ expected false detections, hence the probability of false
detection for each of $n + N_f$ peaks is reduced by a factor of $N_f /
(n + N_f)$. (In other words, we should expect only $N_f$ false detections (noise peaks) for every $n$ actual sources detectable above a given threshold.) CRUSH uses the number of sources $N(>\Sigma)$ that were already
extracted above significance $\Sigma$ minus the expected false detection rate
$N_f(\Sigma)$ as a self-consistent proxy for $n$, which is a reasonable assumption when prior knowledge of the actual underlying counts is not readily available (as in our case). As such, the individual confidence
levels of consecutive detections are estimated as

\begin{equation}
C(\Sigma) = 1 - \frac{ N_f(\Sigma) } { N(>\Sigma) } \left( 1 -
e^{-N_f(\Sigma)} \right).
\label{eq:confidence}
\end{equation}

\subsection{Deboosting}
\label{sec:deboost}
Deboosting is a statistical correction to the observed flux 
densities, when source counts fall steeply with increasing brightness (e.g. \citet{2010ApJ...718..513C} and references therein). 
Thus, in a statistical ensemble of sources, the same observed flux 
arises more often from one of many fainter sources than from the few 
brighter ones, relative to the measured value. 
We assume a measurement with Gaussian noise (validated by the closely Gaussian jackknife noise distribution) and a 2\,mm 
source count model scaled from observationally constrained 850$\mu$m counts (e.g. 
\citet{2006MNRAS.372.1621C}, \citet{2009ApJ...707.1201W}) assuming $T_d/(1+z) \gtrsim$ 10~K 
\citep{2006ApJ...650..592K} and dust 
emissivity index ($\beta$) of 1.5 \citep{2010ApJ...717...29K}. We also deboosted our data using the physical number-count models of \citet{2011ApJ...742...24L} and \citet{2011A&A...529A...4B},  see section \ref{sec:counts}.

For deboosting we followed a Bayesian recipe, such as described in \citet{2005MNRAS.357.1022C, 2006MNRAS.372.1621C}:

\begin{equation}
	p(S_i | S_o, \sigma) \propto p(S_i) ~ p(S_o, \sigma | S_i)
\end{equation}

expressing the probability of intrinsic source flux $S_i$ in terms of the observed flux $S_o$ and its measurement uncertainty $\sigma$.

However, we made some important modifications to the recipe to account for the possibility that the observed flux arises from multiple overlapping galaxies, and we account for confusion. Accordingly, we replace the single isolated source assumption $p(S_i) \propto \frac{dN}{dS}(S_i)$ of \citet{2005MNRAS.357.1022C, 2006MNRAS.372.1621C}, with the compound probability that one or more (up to $m$) resolved sources in the beam contribute to an aggregated intrinsic flux $S_i$:

\begin{equation}
p(S_i) = \int_0^{S_i} dS_1 ~ ... \int_0^{S_{m-1}} dS_m ~ \pi(S_1)~ ... ~\pi(S_m) ~ \delta \left( S_i - \sum_{k=1}^m S_k \right).
\label{eq:intrinsic-density}
\end{equation} 

Inside the integrals is the product of the individual component probability densities $\pi(S_k)$, which correspond to $S_i$ arising from a specific combination of ($S_1 ~ ... ~ S_m$) individual components. The delta function ensures that the component fluxes considered add up to the total intrinsic flux $S_i$ when integrated. Each nested integral for $S(k)$ is performed up to the previous flux $S_{k-1}$, indicating that each successive component $S_k$ is no brighter than the previous one $S_{k-1}$, and ensuring that each particular combination of fluxes is counted one time only. Once the fluxes in the outer integrals sum up to $S_i$, the remaining inner integrals can be skipped (in numerical implementations) corresponding to fewer than $m$ actual contributors (or to keep to a more formal notation we can add the delta function at zero, i.e.\ $\delta(S_k)$, to the definition of $\pi(S_k)$ below to achieve exactly without omitting the any inner integrals). When overlaps are ignored ($m$=1), Eq.~\ref{eq:intrinsic-density} reduces to $p(S_i) = \pi(S_i)$ naturally (no integration required). 

The differential source counts $\frac{dN}{dS}(S_k)$ determine the probability that there is at least one intrinsic source with brightness $S_k$ in the beam, resolved or unresolved. The distinction between resolved and unresolved sources is important: unresolved sources cause a symmetric widening of the map noise distribution (confusion noise) compared to the experimental noise (e.g.\ radiometric down-integration as measured by a jackknife); resolved sources, on the other hand, detected or not, will manifest as an excess of flux on the positive side of the observed flux (or S/N) distribution. Deboosting naturally needs to consider resolved sources only. 

If one distributes $N$ sources randomly in some large area $A$ (for simplicity's sake let's consider the same area to which the counts are normalized, whether deg$^2$ or sr) with $N_b = A / 2 \pi \sigma_b^2$ ($N_b$$\gg$1) independent beams (FWHM$\approx$2.35\,$\sigma_b$), then the chance that none of these sources fall into a given beam (our detection beam) is:

\begin{equation}
q(N) = \left( 1 - \frac {1 } { N_b } \right) ^{N} \approx e^{-N / N_b}
\end{equation}

Therefore, at a given flux density $S_k$ the probability density of a source with brightness $S_k$ being resolved (i.e. unblended with brighter ones) is

\begin{equation}
\pi_+(S_k) \approx \frac {1} {N_b} ~ \frac {dN} {dS} (S_k) ~ e^{-N(>S_k) / N_b},
\end{equation}

in terms of the integrated number counts $N$($>$$S$), and the corresponding differential counts $\frac{dN}{dS}$. Here, $\pi_+(S_k)$ measures the probability density that at least one resolved source with flux $S_k$ falls inside the detection beam, and does not exclude the possibility of further fainter components within the same beam (hence the plus sign as the subscript). Using $\pi_+(S_k)$, however, we can easily express the probability density $\pi(S_k)$ for exactly one component with $S_k$ in a given beam, by simply subtracting the integrated probability that there is at least one other fainter object in that same beam with the first one:

\begin{equation}
\pi(S_k) = \pi_+(S_k) \left(1 - \int_0^{S_k} \pi_+(S)~dS \right).
\end{equation}

For the highest order $m$ under consideration, we may truncate by setting $\pi(S_m) \approx \pi_+(S_m)$. The approximation is valid as long as $m$ is chosen to be large enough such that resolved overlaps with further components (the right-hand integral term) are negligible.

For the particular case of the GISMO 2~mm sources, we considered up to 3 overlapping components ($m$=3) contributing to the observed fluxes. We verified that this was sufficient, as we noticed no measurable degree of incremental change in the deboosted values (and profiles) between $m$=2 and $m$=3. We chose $m$=3 to be on the safe side. At the same time allowing for at least 2 overlapping components instead of just a single isolated source ($m$=1) did have a significant impact on the deboosting results, justifying our modified approach.

Since our source extraction algorithm determines the map zero level as the mode (not the mean) of the distribution, the extracted source fluxes are easily measured against the unresolved background. And, because our deboosting method is based on resolved sources only, it also means that no additional zero-level adjustment is necessary. As a result, the distribution naturally does not extend to negative fluxes, as is demonstrated by the posterior probability distributions of the extracted GISMO sources shown in the appendix.


\subsection{Extraction -- Results}
\label{sec:results}

\begin{deluxetable*}{lllllcl}
\tablecolumns{8}
\tablewidth{0pc}
\tablecaption{GDF 2mm Sources}
\tablehead{
\colhead{ID}  & \colhead{RA}    & \colhead{DEC}   & \colhead{$S$} & \colhead{S/N} & \colhead{confidence}  & \colhead{$N_f$}\\
\colhead{}   & \colhead{[J2000]}    & \colhead{[J2000]}  & \colhead{[mJy]}  & \colhead{} & \colhead{[\%]} & \colhead{}}
\startdata                                                                                  
GDF-2000.1 &   12:36:33.98 & 62:14:08.0 &  0.79$\pm$0.14 &  5.53   & 100  & 0.000  \\
GDF-2000.2 &   12:37:05.95 & 62:11:47.2 &  0.67$\pm$0.16 &  4.18   & 99   & 0.021  \\
GDF-2000.3 &   12:37:12.17 & 62:13:20.4 &  0.78$\pm$0.19 &  4.10   & 99   & 0.029  \\
GDF-2000.4 &   12:36:29.54 & 62:13:11.7 &  0.53$\pm$0.15 &  3.51   & 87   & 0.330  \\ 
GDF-2000.5 &   12:36:51.40 & 62:15:39.1 &  0.61$\pm$0.18 &  3.37   & 87   & 0.541  \\
GDF-2000.6 &   12:36:52.06 & 62:12:26.4 &  0.42$\pm$0.13 &  3.33   & 89   & 0.627  \\
GDF-2000.7 &   12:36:57.09 & 62:13:29.7 &  0.40$\pm$0.12 &  3.24   & 89   & 0.875 \\
GDF-2000.8 &   12:37:10.12 & 62:13:35.5 &  0.54$\pm$0.17 &  3.22   & 89   & 0.918  \\
GDF-2000.9 &   12:36:36.54 & 62:11:13.5 &  0.51$\pm$0.16 &  3.17   & 84   & 1.117  \\
GDF-2000.10 &  12:36:25.16 & 62:14:10.5 &  0.87$\pm$0.29 &  3.06   & 84   & 1.630  \\
GDF-2000.11 &  12:36:45.88 & 62:14:42.2 &  0.43$\pm$0.14 &  3.04   & 84   & 1.735  \\
GDF-2000.12 &  12:36:56.17 & 62:10:19.1 &  0.54$\pm$0.19 &  3.02   & 84   & 1.828 
\enddata
\tablecomments{$N_f$ is the expected cumulative false detection rate  as defined in Section~\ref{sec:stats}. The maps are beam-smoothed (by 17.5'' FWHM) to an effective $24.7''$ FWHM image resolution for
point source extraction}
\label{tab:source_fluxes} 
\end{deluxetable*}

\begin{deluxetable*}{lccc}
\tablecolumns{8}
\tablewidth{0pc}
\tablecaption{Deboosted 2mm Flux Densities Assuming Different Source Counts or Models}
\tablehead{
\colhead{ID}  & \multicolumn{3}{c}{deboosted $S'$ [mJy]} \\
\cline{2-4}
& \colhead{Lapi et al.} & \colhead{Coppin et al.}& \colhead{Bethermin et al.} }
\startdata                                                                                  
GDF-2000.1 &       0.75  $\pm$  0.16 &      0.69 $\pm$  0.14 &      0.71 $\pm$  0.16 \\ 
GDF-2000.2  &      0.56 $\pm$  0.21 &      0.53 $\pm$  0.18 &      0.51 $\pm$  0.20 \\
GDF-2000.3  &      0.63 $\pm$  0.25 &      0.59 $\pm$  0.22 &      0.56 $\pm$  0.25 \\
GDF-2000.4  &      0.39 $\pm$  0.20 &      0.37 $\pm$  0.18 &      0.35 $\pm$  0.19 \\
GDF-2000.5  &      0.40 $\pm$  0.24 &      0.38 $\pm$  0.21 &      0.34 $\pm$  0.23 \\
GDF-2000.6  &      0.31 $\pm$  0.16 &      0.29 $\pm$  0.15 &      0.28 $\pm$  0.16 \\
GDF-2000.7  &      0.29 $\pm$  0.15 &      0.27 $\pm$  0.14 &      0.26 $\pm$  0.15 \\
GDF-2000.8  &      0.35 $\pm$  0.22 &      0.33 $\pm$  0.20 &      0.30 $\pm$  0.20 \\
GDF-2000.9  &      0.33 $\pm$ 0.20 &      0.31 $\pm$  0.19 &      0.28 $\pm$  0.19 \\
GDF-2000.10 &      0.40 $\pm$  0.33 &      0.37 $\pm$  0.30 &      0.30 $\pm$  0.29 \\
GDF-2000.11 &      0.28 $\pm $ 0.17 &      0.26 $\pm$ 0.16 &      0.25 $\pm$  0.17 \\
GDF-2000.12 &      0.33 $\pm$  0.23 &      0.31 $\pm$  0.21 &      0.24 $\pm$  0.20
\enddata
\tablecomments{Derived by extrapolation of the \cite{2006MNRAS.372.1621C} SHADES number counts to 2~mm counts, or from the 
\cite{2011ApJ...742...24L} or \cite{2011A&A...529A...4B} models.}
\label{tab:deboost}
\end{deluxetable*}

\begin{deluxetable*}{lllllllcl}
\tablecolumns{9}
\tablewidth{0pc}
\tablecaption{GDF 2~mm Sources with (sub-)mm counterparts}
\tablehead{
\colhead{ID}    & \colhead{$z$} & \colhead{ID}  & \colhead{$S$($850\mu$m)}  & \colhead{$S'$($850\mu$m)}   & \colhead{ID}  & \colhead{$S$(1.2\,mm)} & \colhead{$S'$(1.2\,mm)} & \colhead{$\delta r$}      \\
\colhead{}       & \colhead{} & \colhead{} & \colhead{[mJy]}  & \colhead{[mJy]}  & \colhead{} & \colhead{[mJy]} & \colhead{[mJy]}  &\colhead{[$''$]}}
\startdata            
GDF-2000.1 & 4.042 & GN 850.10 & 11.3$\pm$1.6 &  8.6$\pm$4.8 & AzGN03  & 5.2$\pm$0.6 & 4.7$\pm$1.7 & 6.0 \\ 
GDF-2000.3  & 1.992 & GN 850.39  & 7.4$\pm$2.0 & 3.8$\pm$2.8  &  AzGN07A$^*$  & blend$^*$ & \nodata  & 9.8 \\
                      &           &     &  & & GN1200.3 & 3.9$\pm$0.6 & 3.3$\pm$1.3 \\
GDF-2000.6  & 5.183 & GN 850.14 & 5.9$\pm$0.3  & 5.9$\pm$0.3  & AzGN14   & 2.9$\pm$0.6 & 2.2$\pm$1.0  & 1.6 \\
GDF-2000.8  & 1.97 & &  $<6$  &    \nodata & AzGN07B$^*$  & blend$^*$ & \nodata  & 13.8 \\
GDF-2000.11  & 2.30 & GN 850.12 & 8.6$\pm$1.4  & 6.4$\pm$3.6 & AzGN08  &  3.0$\pm$0.6 & 2.4$\pm$1.1 & 6.6
\enddata
\tablecomments{MAMBO/AzTEC 1.2~mm data are from \citet{2008MNRAS.389.1489G} and \citet{2008MNRAS.391.1227P}, SCUBA $850 \mu$m data from \citet{2003MNRAS.344..385B}, \citet{2005MNRAS.358..149P}, 
 and \citet{2008MNRAS.389.1489G}. $z$ is the measured redshift. The deboosted flux values $S'$ are based on the SHADES counts, using the same equations used for calculating the 2~mm data counts in order to be consistent with the deboosting of the 2~mm fluxes shown in Table \ref{tab:deboost}.\\ 
$^*$ Both, GDF-2000.3 and GDF-2000.8, are associated with AzGN07, which implies that this 1.2~mm source is a blend of two sources. 
}
\label{tab:counterparts}
\end{deluxetable*}


\begin{deluxetable*}{lccccccccc}
\tablecolumns{11} 
\tablewidth{0pc}
\tablecaption{low S/N GDF 2mm Sources -- identified through (sub-)mm counterparts}
\tablehead{
\colhead{ID}  & \colhead{RA}    & \colhead{DEC} & \colhead{$S$(2mm)} & \colhead{$z$} & \colhead{ID}  & \colhead{$\lambda$} & \colhead{$S$}  & \colhead{$S'$} & \colhead{$\delta r$}  \\
\colhead{} & \colhead{[J2000]}    & \colhead{[J2000]}  & \colhead{[mJy]}   & \colhead{}    &  \colhead{}  & \colhead{} & \colhead{[mJy]}  & \colhead{[mJy]} & \colhead{[$''$]} }
\startdata        
GDF-2000.13 & 12:37:12.01 & 62:12:14.0 &  0.52$\pm$0.20 &  2.91    & GN 850.21  & 850 $\mu$m & 5.7$\pm$1.2 & 3.8$\pm$2.3 & 8.1 \\ 
            &             &            &                &  2.91    & GN 1200.29 & 1200 mm    & 2.6$\pm$0.6 & 1.9$\pm$0.9 & 0.0 \\ 
GDF-2000.14 & 12:36:27.40 & 62:12:13.8 &  0.50$\pm$0.19 &  4.69    & AzGN10     & 1200 mm    & 2.6$\pm$0.6 & 1.9$\pm$0.9 & 4.9 \\
GDF-2000.15 & 12:36:45.00 & 62:11:47.1 &  0.36$\pm$0.15 &  \nodata & GN 850.28  & 850 $\mu$m & 1.7$\pm$0.4 & 0.6$\pm$0.6 & 1.8
\enddata
\tablecomments{$z$ is the measured redshift. Deboosted flux values $S'$ were calculated directly (850$\mu$m) from or extrapolated (1.2~mm) from the SHADES number counts.}
\label{tab:low_sn}
\end{deluxetable*}

\begin{deluxetable*}{lccc}
\tablecolumns{8}
\tablewidth{0pc}
\tablecaption{Deboosted 2mm Flux Densities Assuming Different Source Counts or Models}
\tablehead{
\colhead{ID}  & \multicolumn{3}{c}{deboosted $S'$ [mJy]} \\
\cline{2-4}
 & \colhead{Coppin et al.} & \colhead{Lapi et al.}& \colhead{Bethermin et al.} }
\startdata                                                                                  
GDF-2000.13 &      0.26 $\pm$  0.21 &      0.24 $\pm$  0.20 &      0.21 $\pm$  0.19 \\
GDF-2000.14 &      0.25 $\pm$  0.20 &      0.23 $\pm$  0.19 &      0.20 $\pm$ 0.18 \\
GDF-2000.15 &      0.21 $\pm$  0.15 &      0.17 $\pm$  0.14 &      0.17 $\pm$  0.14 
\enddata
\tablecomments{Derived by extrapolation of the \cite{2006MNRAS.372.1621C} SHADES number counts to 2~mm counts, and from the 
\cite{2011ApJ...742...24L} or \cite{2011A&A...529A...4B} models.}
\label{tab:low_sn_deboost}
\end{deluxetable*}

\begin{deluxetable*}{llllllll}
\tablecolumns{8}
\tablewidth{0pc}
\tablecaption{GDF 2mm Sources -- derived parameters \label{tab:singleT}}
\tablehead{
\colhead{ID}  & \colhead{$|\hat{\chi}^2|$} & \colhead{$T_c$} & \colhead{$\log M_d$} & \colhead{$\log L$}  &\colhead{$q_L$} & \colhead{$q_{IR}$} & \colhead{$\tau_{\rm peak}$} \\ 
 \colhead{} & \colhead{}  & \colhead{[K]} & \colhead{[$M_\odot$]} & \colhead{[$L_\odot$]} & \colhead{} & \colhead{} & \colhead{}}
\startdata
GDF 2000.1   & 0.70  & 51.2 $\pm$ 2.0 & 8.53 $\pm$ 0.07 & 13.52 $\pm$ 0.06 & 2.83$\pm$0.21 & 3.08 $\pm$ 0.21 & 1.28\\
GDF 2000.3  & 1.08 & 40.8 $\pm$ 0.7 & 8.59 $\pm$ 0.14 & 13.10 $\pm$ 0.03  & 2.68 $\pm$ 0.10 & 2.91 $\pm$ 0.10 & 1.08 
\enddata
\tablecomments{All quantities were fitted using the measured redshifts (Table~2),  temperature-distribution model ($dM_d/dT \propto T^{-7.2}$) with $\beta=1.5$ and assuming a 2\,kpc emission diameter. The dust masses assume $\kappa(850\mu$m)=0.15\,m$^2$\,kg$^{-1}$. Uncertainties are 1\,$\sigma$ total errors of the fits to data, which do not include the uncertainties in the redshift values. The following quantities are shown in the table:  $|\hat{\chi}^2|$ residual scatter around the fit, $T_c$ temperature of the dominant cold component, $M_d$ dust mass, $L$ integrated IR luminosity, $q_L$ radio--(F)IR correlation constant as defined in \citet{2010ApJ...717...29K}, $q_{\rm IR}$ radio--(F)IR correlation constant as defined in \citet{2010MNRAS.402..245I}, and  $\tau_{\rm peak}$ optical depth  around the IR emission peak.}
\label{tab:source_prop}
\end{deluxetable*}

\begin{figure}
\includegraphics[width=3.3in]{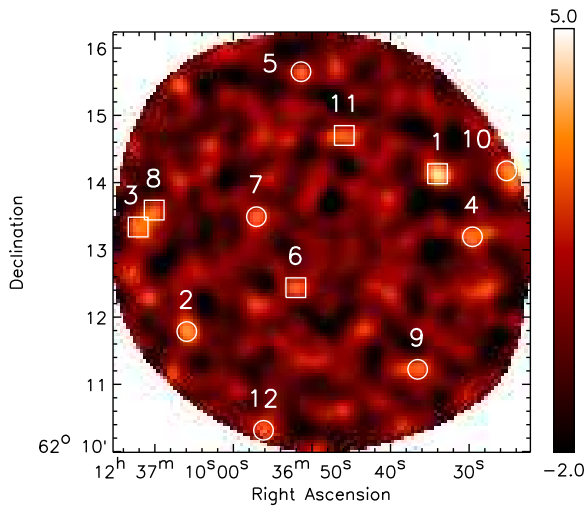}
\caption{Map of the S/N for the smoothed and filtered GISMO Deep Field (GDF).
The applied point source filter yields an effective image resolution of $24.7''$ FWHM. Sources at S/N $>3$ (Table \ref{tab:source_fluxes}) are marked with circles
(sized as the $17.5''$ FWHM of the diffraction limited GISMO beam) 
and squares (for sources with 1.2~mm and/or 850~$\mu$m counterparts).\label{GDF}}
\end{figure}

\begin{figure}
\includegraphics[width=3.3in]{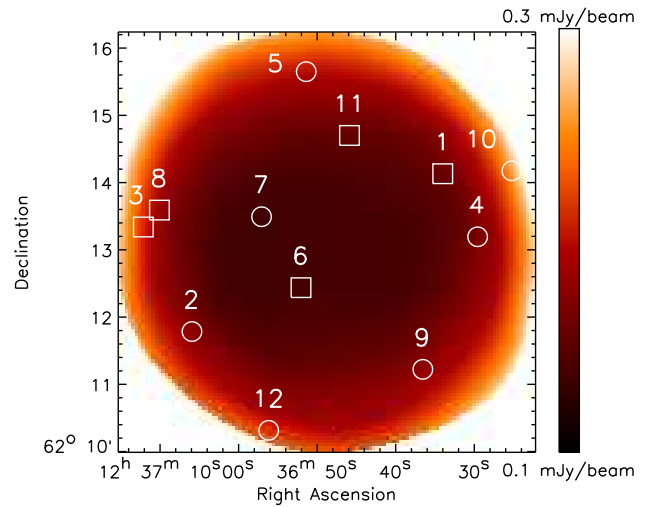}
\caption{Noise map for the smoothed and filtered GISMO Deep Field (GDF). The
 noise in the innermost $\sim 4'$ diameter in the map is $\sim 135 \mu$Jy/beam. 
Sources at S/N $>3$ (Table 1) are marked with circles
(sized as the $17.5''$ FWHM of the diffraction limited GISMO beam) 
and squares (for sources with 1.2mm and/or 850 $\mu$m counterparts).
\label{GDF-rms}}
\end{figure}

Figure \ref{GDF} shows the beam-smoothed signal-to-noise map of the  2\,mm GDF.  Figure \ref{GDF-rms} shows the noise map, demonstrating that the innermost $4'$ of the GDF observations have an rms of between 130 and 140 $\mu$Jy. The source extraction was performed out to twice this level, i.e. up to  $\sim$260$ \mu$Jy rms (Fig. \ref{GDF-rms}). The  area for the source extraction is $\sim$31 square arcminutes ($\sim$321 GISMO beams). 

The source extraction algorithm finds 12 positive sources and 3 negative ``sources''. The number of significant negatives is consistent with the expected false detection rate of $N_f = 2$ for a 2.99\,$\sigma$ detection threshold in the extraction area (see Section \ref{sec:stats}).
The measured 2~mm fluxes range from 400 to 870  $\mu$Jy.

Table~\ref{tab:source_fluxes} shows the measured fluxes of the 12 positive sources with the achieved signal-to-noise, the estimated detection confidence levels, and the expected cumulative false detection rate $N_f$.  The fluxes presented in the table show the measured fluxes, while Table \ref{tab:deboost} shows 
the de-boosted values, using a scaled version of the broken power-law SHADES galaxy number 
counts \citep{2006MNRAS.372.1621C}, the \cite{2011ApJ...742...24L} number counts, as well the 
values corresponding to the \cite{2011A&A...529A...4B} counts.

In order to identify (sub)millimeter counterparts of our sources we cross correlated our data with the 1.16~mm combined MAMBO/AzTEC source catalog containing the 1.1~mm the AzTEC data from \citet{2008MNRAS.391.1227P}, and the MAMBO source catalog from \citet{2008MNRAS.389.1489G}, as well as the SCUBA $850 \mu$m source catalog of GOODS-N (\citet{2003MNRAS.344..385B}, as well as \citet{2004MNRAS.355..485B}, 
\citet{2005MNRAS.358..149P}, and \citet{2006MNRAS.370.1185P}).
Table \ref{tab:counterparts} shows the measured and deboosted 1.2~mm and $850~\mu$m fluxes  of the sources with counterparts, if available. The equation we used for the maximum allowable separation between the GISMO/MAMBO/AzTEC/SCUBA sources in order to be considered a counterpart is given by:
\begin{equation}
 r_{max}^2 = 4  \sigma_{p}^2 - 2  \sigma_{beam}^2  \rm{ln} \left( 1 - \frac{2}{SNR} \right),
 \label{search-r}
 \end{equation} 
 
 where $r_{max}$ is the search radius where the counterpart must fall with $> 98\%$ confidence, $ \sigma_{p}$ is the 1 sigma catalog position error combined with the rms astrometric accuracy of our map ($<$0.8'') in quadrature,  $\sigma_{beam}$ is the 1 sigma beam size $\sim FWHM/2.35$ , and SNR is the observed signal-to-noise ratio of the detection. When searching for a counterpart in another low S/N dataset of (sub-)mm data set, a second identical SNR-dependent term needs to be added to the search-radius expression above, reflecting the inherent positional uncertainty of the other known mm-wave detection. In section \ref{sec:sims}, we demonstrate the applicability of equation \ref{search-r} for our dataset.

\begin{figure}
\includegraphics[angle=270,width=3.3in]{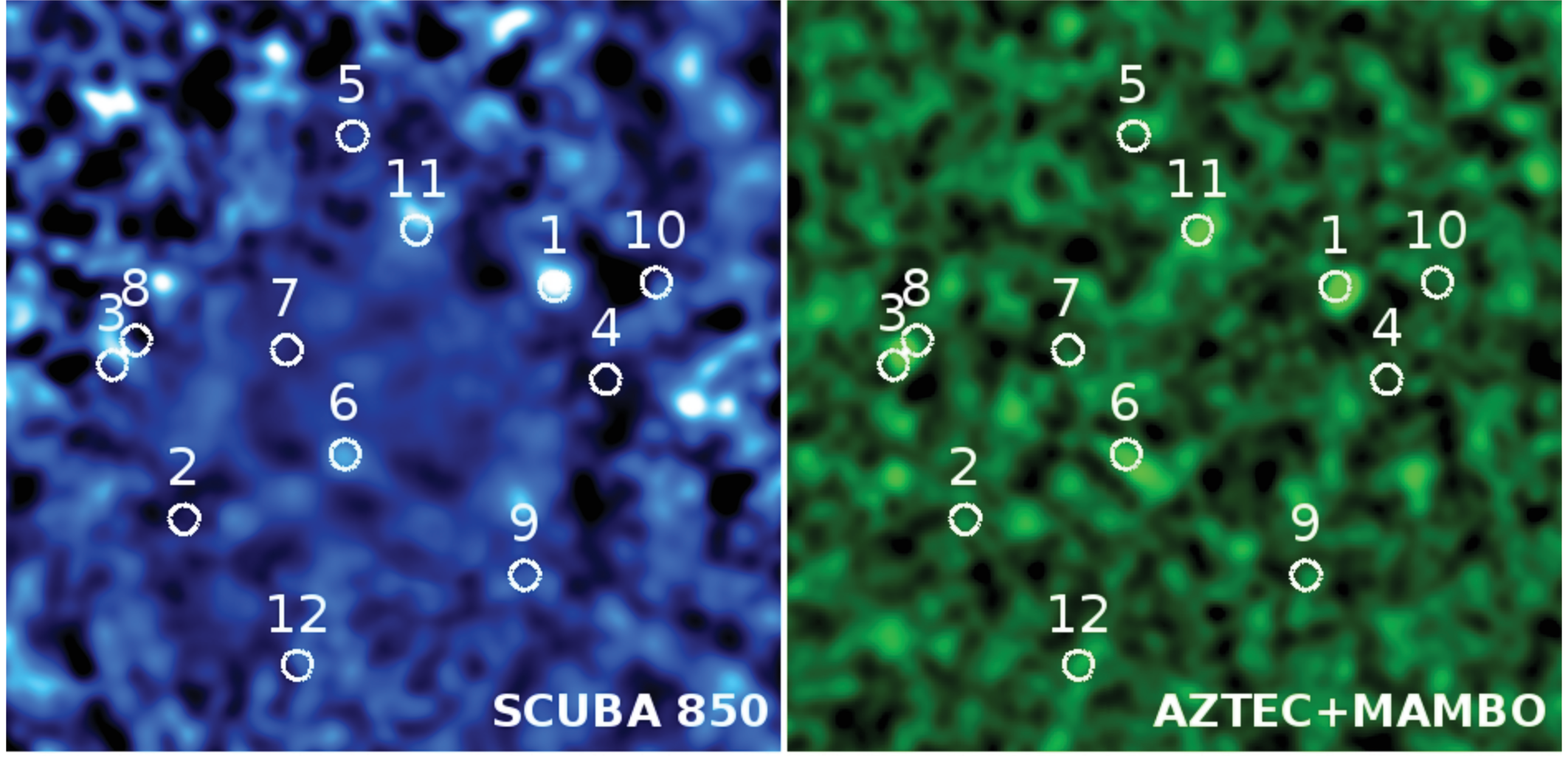}
\includegraphics[angle=270,width=3.3in]{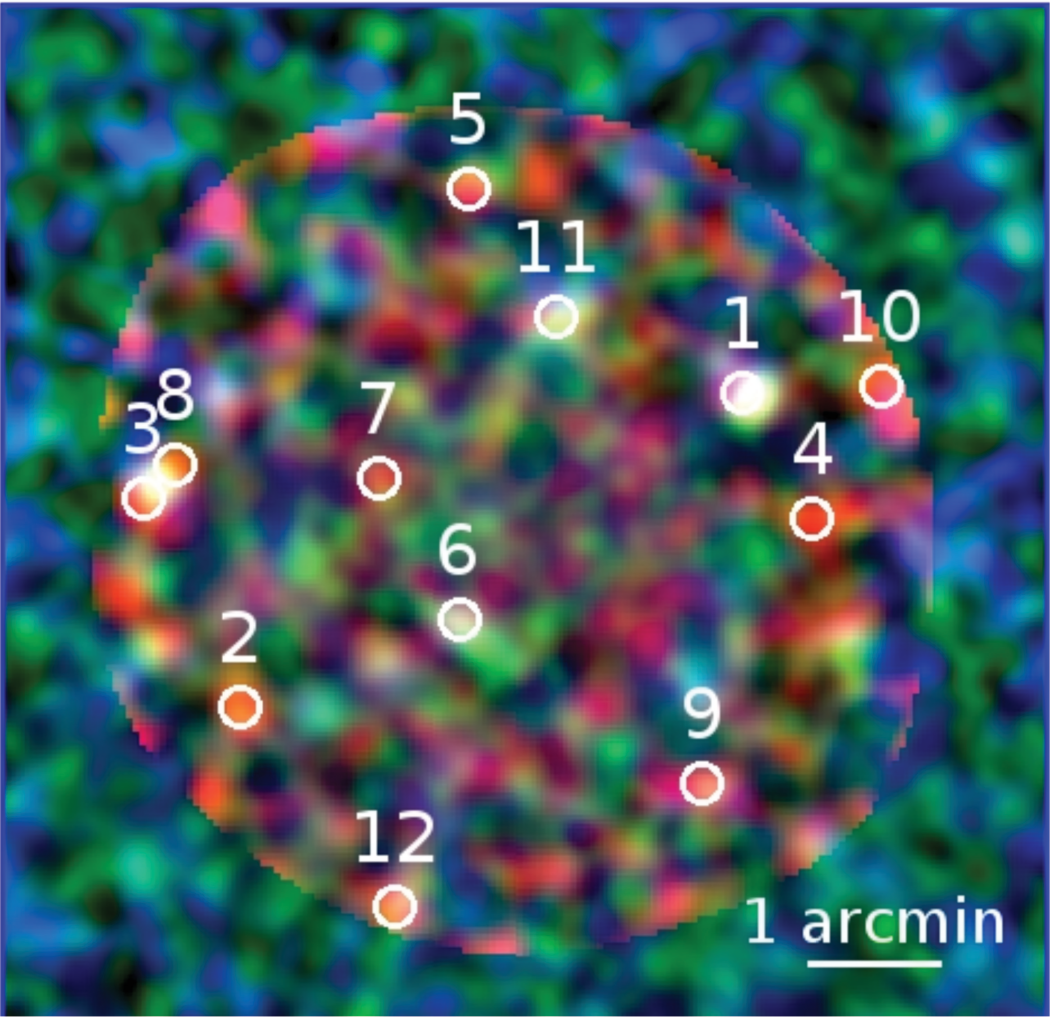}
\caption{The GISMO Deep Field Sources at multiple wavelengths. The SCUBA 850 $\mu$m map \citep{2005MNRAS.358..149P} 
is shown in blue. The AzTEC+MAMBO 1.2 mm map \citep{2011MNRAS.410.2749P}, is shown in green. The 2 mm GISMO image
is shown in red.\label{GDF-RGB}}
\end{figure}

Combining the GISMO, MAMBO/AzTEC/SCUBA observations (Fig. \ref{GDF-RGB}), we identify three additional sources with detection confidence level of $> 80\%$. These are tabulated in Table \ref{tab:low_sn} and \ref{tab:low_sn_deboost}.
The median and mean redshifts of all sources with counterparts and known redshifts are $\tilde{z} = 2.91 \pm 0.94$ and $\bar{z} = 3.3$, respectively.

\subsection{GDF sources without a counterpart}
Seven of the detected 2\,mm sources have no counterpart in either the MAMBO/AzTEC or the SCUBA data (GDF-2000.2, GDF-2000.4, GDF-2000.5, GDF-2000.7, GDF-2000.9, GDF-2000.10, and GDF-2000.12). Considering their observed signal to noise ratio and the three negative detections in the map, plus the statistical expectation of 2 false detections as derived in section \ref{sec:stats}, the most likely scenario is that 5 of these are real detections. 

\begin{figure}
\includegraphics[angle=0,width=3.3in]{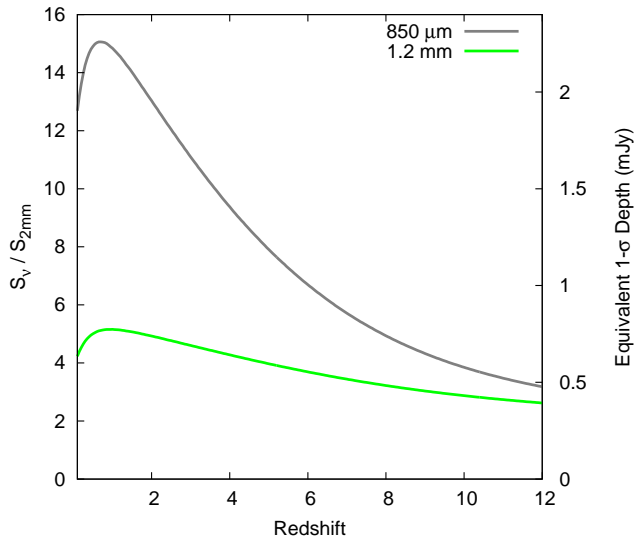}
\caption{Redshift-dependent GDF colors at 850~$\mu$m and 1.2~mm, relative to the 2~mm fluxes, based on the same SED model as Fig.~\ref{K-correct}. On the right axis, we indicate the typical $1-\sigma$ depths that an 850~$\mu$m or 1.2~mm map would have to reach, as a function of redshift, to provide equivalent coverage of the sources to that of the GISMO deep field in this paper.}
\label{rms-color}
\end{figure}

Figure \ref{rms-color} shows the redshift-dependent GDF equivalent map sensitivities at 850~$\mu$m and 1.2~mm for the detection of a galaxy with the SED shown in Fig. \ref{K-correct}. The figure demonstrates that the equivalent 1.2~mm source sensitivity requirement to match the GDF source sensitivity varies by about a factor of two with redshift, whereas the required depth of the  850~$\mu$m SCUBA map varies significantly. In order to achieve the same GDF source sensitivity the $1~\sigma$ map noise rms at 1.2~mm would need to be $\sim 0.7$ mJy/beam if the galaxy were at a redshift of 2, while it would require $\sim$0.4 mJy/beam if the same source were at an extremely high redshift. At 850 ~$\mu$m the matching map flux sensitivity requirement ranges from about 2 mJy/beam for a source at a redshift $z\sim2$ to  $\sim0.5$ mJy/beam for a source at extremely high redshifts. Table \ref{tab:nocountrms} shows the actual 850~$\mu$m and 1.2~mm map sensitivities at the seven positions of GDF sources without counterparts.  The table shows that the depth of the 1.2~mm map is quite homogeneous for all sources without counterparts with an rms of about 0.56 mJy for each of those. This means that for our template SED, the source sensitivity of the GDF map exceeds that of the 1.2~mm data for redshifts of about $z=6$ and greater. The situation for the SCUBA data is different, since the sensitivity over that map varies very significantly as is demonstrated by the range as shown in the same table. Only at the position of GDF-2000.7 the sensitivity of the SCUBA map is below an rms of 1 mJy/beam (Table \ref{tab:nocountrms}). For the other sources the 850~$\mu$m  map sensitivities are between 1 and 2 mJy/beam rms. A comparison of these numbers with the equivalent 850~$\mu$m   GDF point source sensitivities shown in Fig. \ref{rms-color} shows that the 2~mm point source sensitivity of all GDF sources without counterparts, with the exception  of GDF-2000.7, exceeds the point source sensitivity of the SCUBA map for redshifts $z > 6$. Taken together, the 850~$\mu$m and 1.2~mm data on non-detected GDF sources are entirely consistent with sources at $z > 6$, assuming the template from Fig. \ref{K-correct} applies.   
However, another aspect when considering that a detected 2~mm source has no counterpart in another catalog is that of completeness or the probability of a low S/N source to be extracted.  For the 850~$\mu$m SCUBA map, e.g., only the completeness of sources with $S/N \gtrsim 5$ is $>  90\%$
 \citep{2006MNRAS.372.1621C}, i.e. the S/N dependent probabilities are similar to those derived in the completeness analysis for the GDF data presented in section \ref{sec:sims}. The situation is similar for the 1.2~mm data. As a result, the S/N limitations  of the  850~$\mu$m SCUBA and 1.2~mm MAMBO/AzTEC data sets combined with the low number of 2~mm sources prevents us from deriving significant redshift constraints on the GISMO sources without counterparts.  Larger deep 2-mm surveys will be needed to photometrically study what fraction of the 2-mm population is at very high-redshift ($z > 6$), one of the main science goals of GISMO. 

\begin{deluxetable*}{lll}
\tablecolumns{3}
\tablewidth{0pc}
\tablecaption{850~$\mu$m and 1.2~mm sensitivities at the positions of the GDF 2~mmm sources without counterparts}
\tablehead{
\colhead{ID}   &  \colhead{$rms$($850\mu$m)}  & \colhead{$rms$(1.2\,mm)}      \\
\colhead{} & \colhead{[mJy/beam]}  & \colhead{[mJy/beam]}  }
\startdata            
GDF-2000.2 & 1.02 & 0.56 \\ 
GDF-2000.4  & 1.54 & 0.57 \\
GDF-2000.5  & 1.64 & 0.55 \\
GDF-2000.7  & 0.42 & 0.55 \\
GDF-2000.9  & 1.09 & 0.56 \\
GDF-2000.10  & 2.04 & 0.56  \\
GDF-2000.12 & 1.85 & 0.55
\enddata
\label{tab:nocountrms}
\end{deluxetable*}


\subsection{Simulations}
\label{sec:sims}

\begin{figure} [t]
   \centering
   \includegraphics[width=3.in]{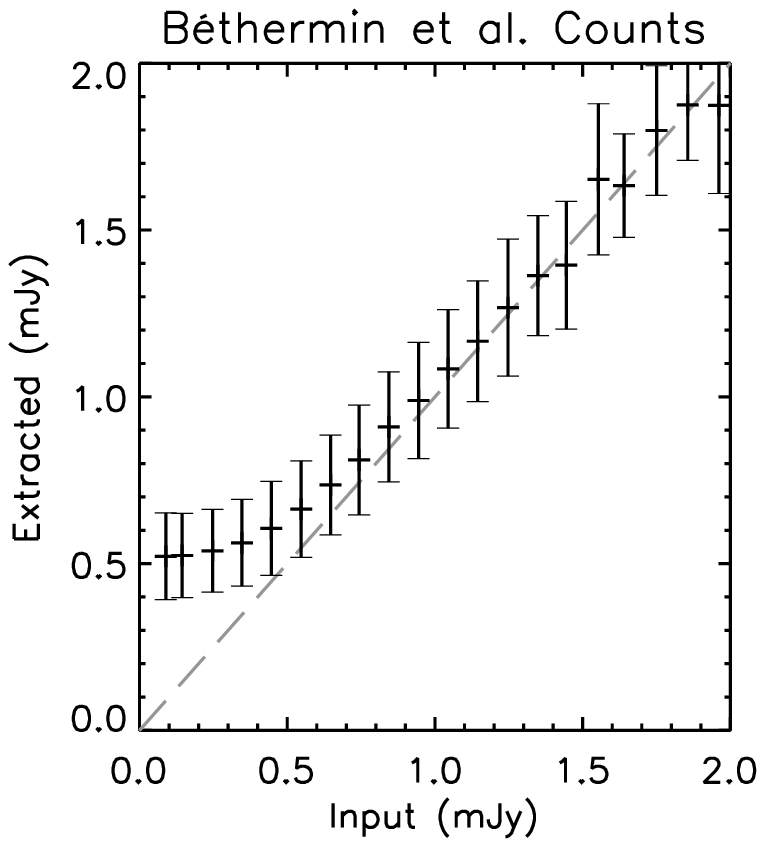} 
   \includegraphics[width=3.in]{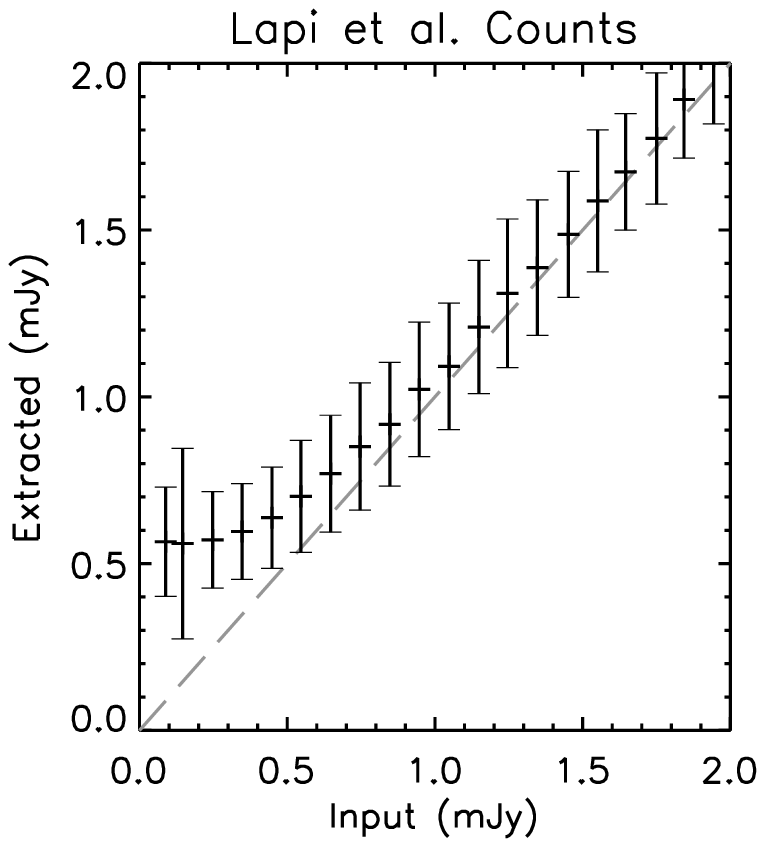}\\
   \caption{Plots of the extracted source flux density as a function 
   of the input flux density for simulations using \citet{2011A&A...529A...4B}
   source counts (top) and extrapolated \citet{2011ApJ...742...24L} source counts
   (bottom). The error bars reflect the $1\sigma$ 
   dispersion for flux density intervals populated by more than 1 
   simulated source. The boosting of the flux densities is evident for 
   sources at $<1$~mJy.
   \label{fig:S-vs-S}}
\end{figure}

\begin{figure}[t] 
   \centering
   \includegraphics[width=3.in]{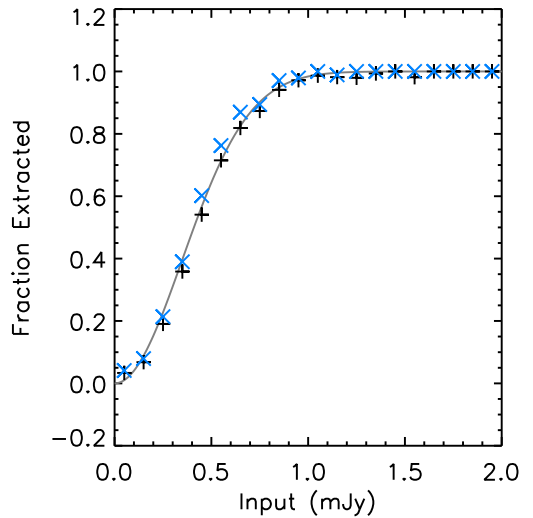} 
   \caption{The completeness (fraction of simulated input sources that are 
   recovered by the source extraction) is plotted as a function of the input
   flux density of the simulated sources. Simulations using \citet{2011A&A...529A...4B} number counts and extrapolated \citet{2011ApJ...742...24L} counts 
   are shown as blue and black respectively. In both cases the completeness drops 
   to 50\% at 0.4 mJy.
   The completeness can be empirically fit by the inverted gaussian function 
   $1-\exp{(-0.5(S/\sigma)^2)}$ where $\sigma = 0.348$ mJy (gray line).\\
   \label{fig:completeness}}
\end{figure}

\begin{figure} 
   \centering
   \includegraphics[width=3.in]{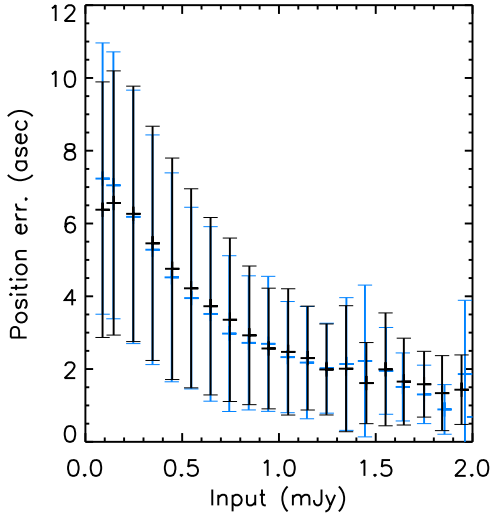} \\
   \caption{The mean positional  errors with $1-\sigma$ error bars of extracted sources are plotted as 
   a function of the input source brightness. The simulations use the \citet{2011A&A...529A...4B} number count model and the extrapolated \citet{2011ApJ...742...24L} model , shown as blue and black respectively. The mean positional errors 
   are $<5.1''$ for sources brighter than 0.4 mJy. 
   \label{fig:position}}
\end{figure}

\begin{figure} 
   \centering
   \includegraphics[width=3.in]{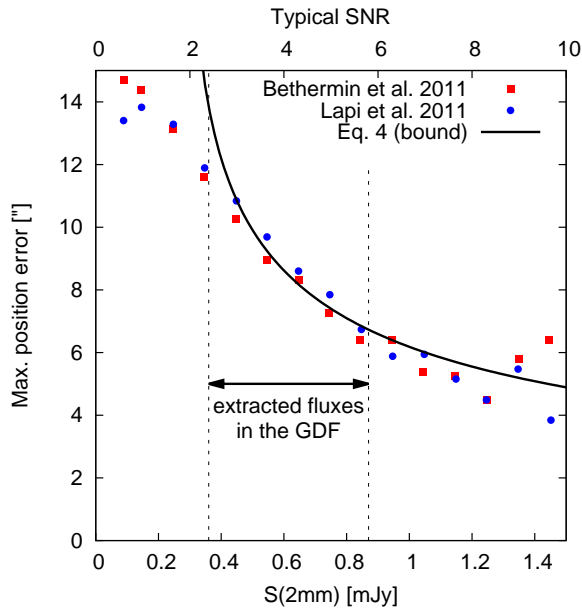}
   \caption{Determining search radii: 2-$\sigma$ position errors of sources extracted from the simulations as a function of input flux (points), vs the prediction from Eq.  \ref{search-r} as a function of S/N (solid curve). To convert the simulated input fluxes to S/N, we estimated an average depth of 0.15 mJy/beam in our map, and we used $\sigma_p = 0$ since there were no intrinsic pointing errors in the simulated data. The dotted lines indicate the range of fluxes extracted from our map, i.e. the range for which search radii are calculated. We note that at low S/N the simulations tend to falsely identify a nearer chance peak (one of the many faint sources filling the map, or a noise peak) as the counterpart to the input. As such, the simulations tend to underestimate the true position errors at low S/N, explaining the deviation from the curve below detection level. Also, the asymptotic behavior at high S/N ($>$10) is not well modeled by Eq. \ref{search-r}, however for the range of fluxes considered, the calculated search radii are in remarkable agreement with simulations, thus justifying our reliance on them.
      \label{fig:position-eq-vs-sim}}
\end{figure}

\begin{figure} 
   \centering
   \includegraphics[width=3.in]{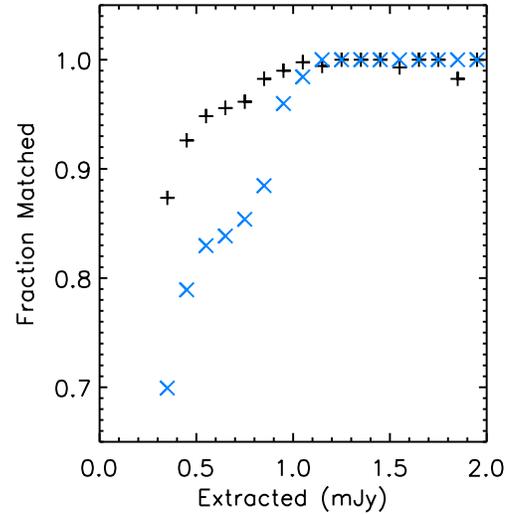} 
   \caption{The reliability (fraction of extracted sources that match input
   simulated sources) is plotted as a function of the input
   flux density of the simulated sources. The simulations use the \citet{2011A&A...529A...4B} number count model and the extrapolated \citet{2011ApJ...742...24L} model , shown as blue and black respectively.\\
   \label{fig:reliability}}
\end{figure}

\begin{figure} 
   \centering
   \includegraphics[width=3.in]{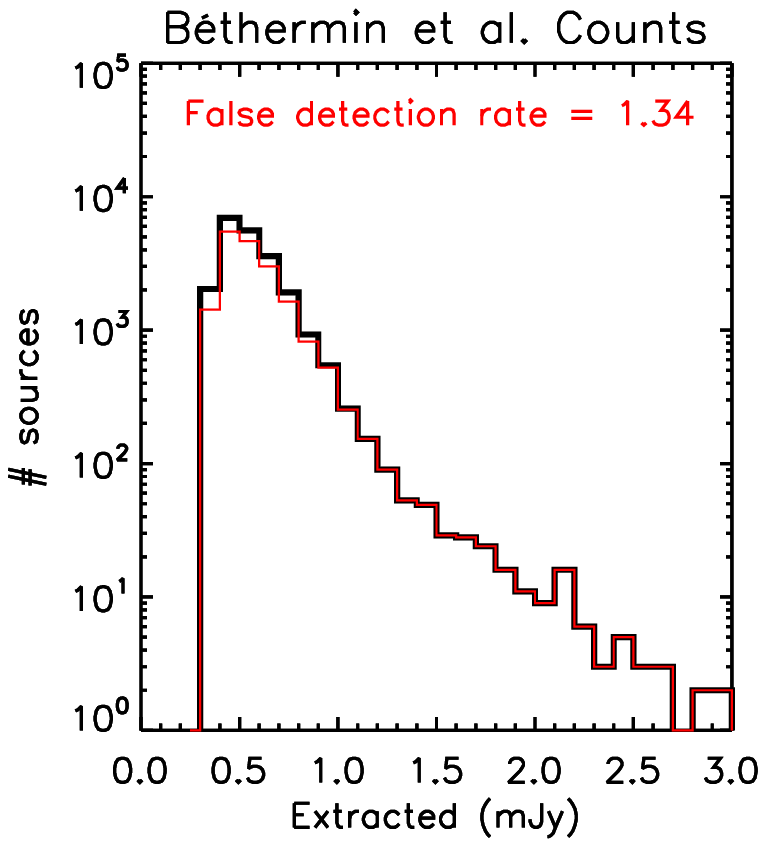} 
   \includegraphics[width=3.in]{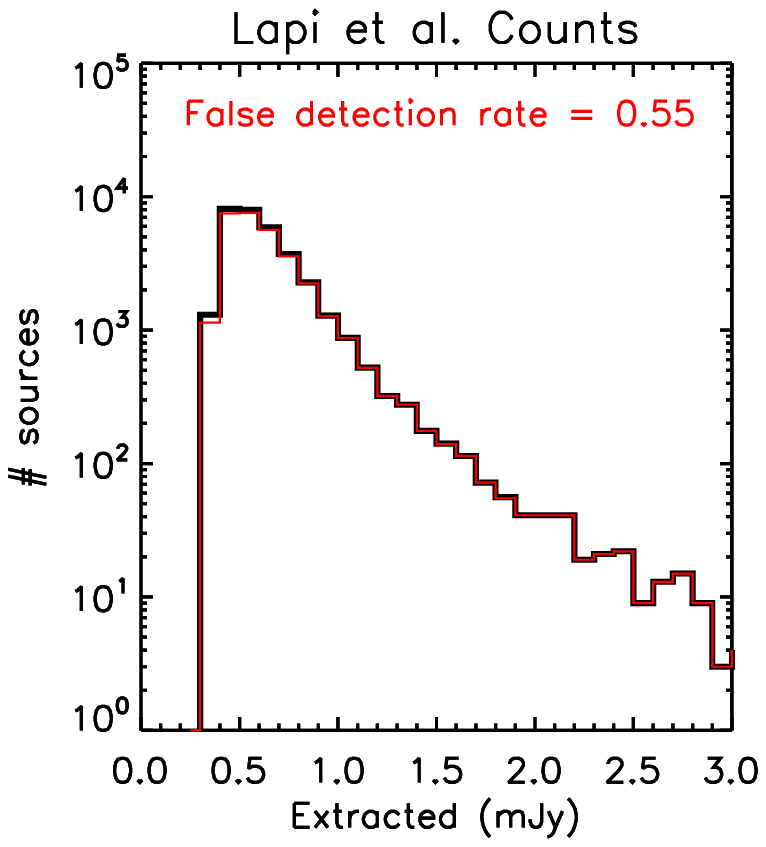}\\
   \caption{Histograms of the brightnesses of the extracted sources (black) and the simulated 
   sources (red). False sources are indicated by an excess number of extracted sources. 
   The mean number of false sources extracted is $\sim1$, but the value is inversely related to the 
   number density of the simulated sources.
   \label{fig:false}}
\end{figure}

\begin{figure} 
   \centering
   \includegraphics[width=3.in]{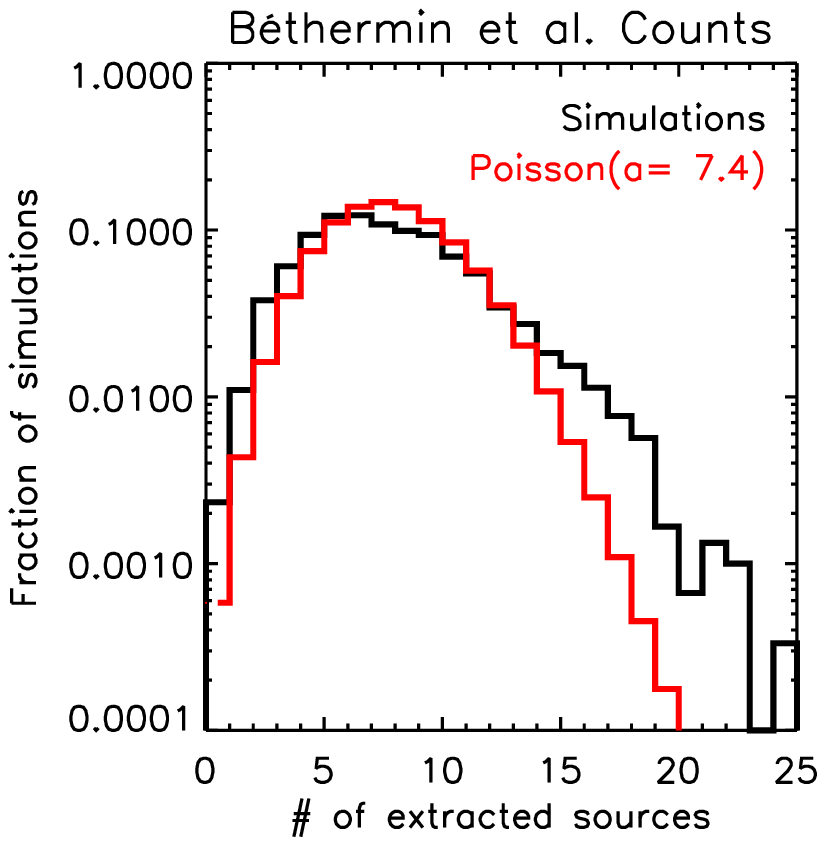} 
   \includegraphics[width=3.in]{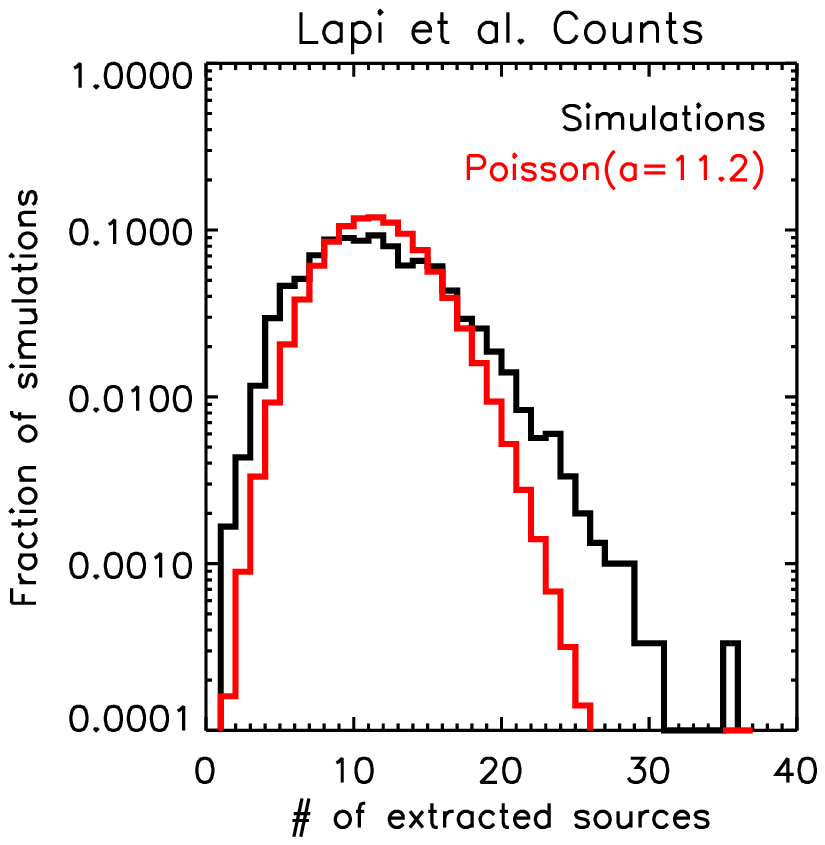}\\
  \caption{Histograms of the number of sources extracted per field, over the set of 
   3000 simulated fields. Overplotted is the Poisson distribution having the same mean number of
   extracted sources.
    \label{fig:simhistsa}}
    \end{figure}
    
\begin{figure}
   \includegraphics[width=3.in]{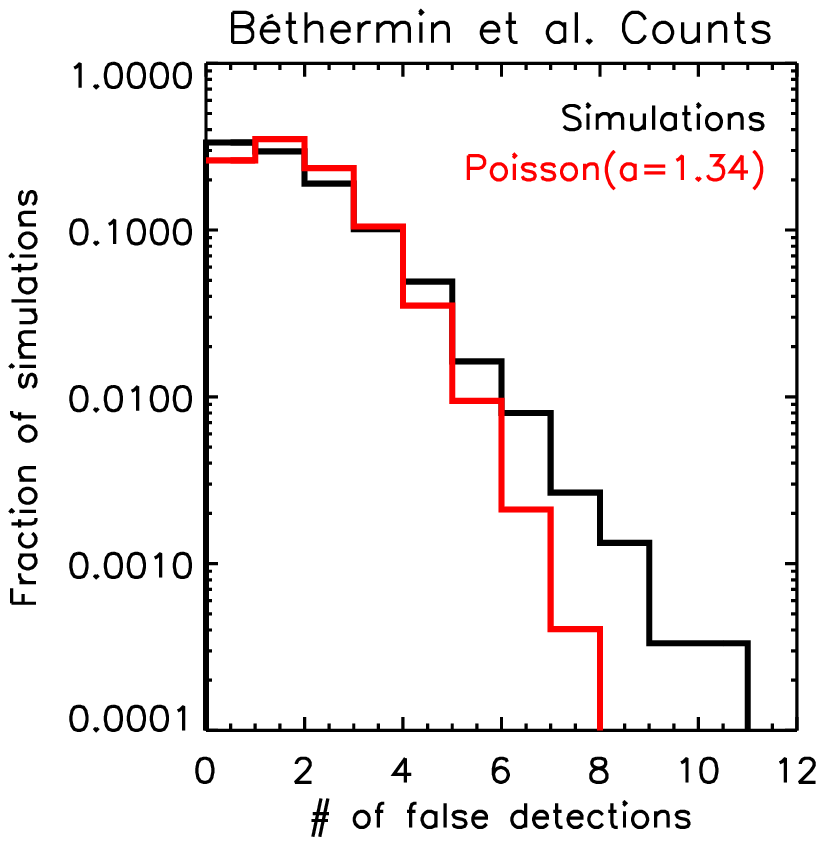} 
   \includegraphics[width=3.in]{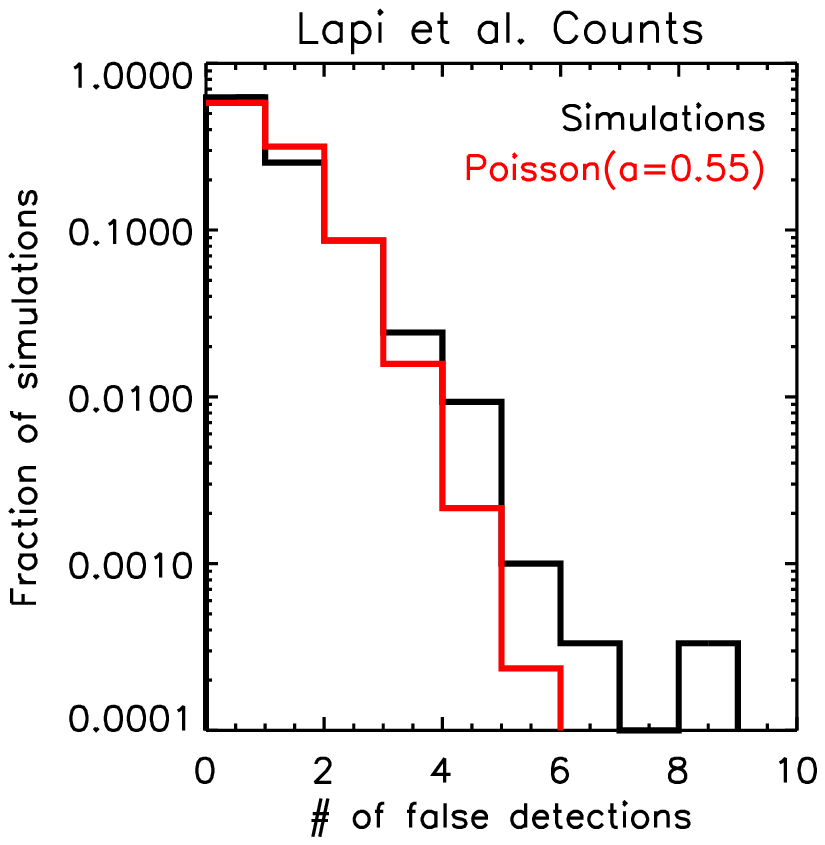}\\
   \caption{Histograms of the number of {\it false} sources extracted per field, 
   over the set of 3000 simulated fields. Overplotted is the Poisson distribution having the 
   same mean number of false sources. 
   \label{fig:simhistsb}}
\end{figure}

Additional characterizations of the extracted sources were obtained through
simulations, generally following the analysis of \citet{2009ApJ...707.1201W}. These simulations
demonstrate also the validity of the equations used for  source extraction   
presented in in sections \ref{sec:stats} and \ref{sec:results}.
The simulations started with the construction of 100 variations
of jackknifed noise maps that were generated from the original data.
These maps provide accurate representations of the noise of the observations. 
We next constructed 3000 versions of the point source distribution across the 
full area of the map (not just the low noise, high coverage regions)
based on the \citet{2011A&A...529A...4B} model and the smoothed and filtered 
GISMO beam. The simulations included sources down to 0.01 mJy in order to 
provide an appropriate confusion background. 
Source brightnesses were chosen at random from the cumulative source 
counts, $N(>S)$, producing a sample size appropriate for 3000 images. 
These sources were then distributed randomly across the full set 
of 3000 noise-free images. Therefore, the peak brightness and the total 
number of sources in each simulated image are subject to variation due to Poisson statistics.
There were an average of 2264 sources in each simulated image.
The jackknife noise maps were added to the simulations, reusing each of the 
noise maps 30 times. The 3000 simulated images were then run through the 
source detection procedure, using the same settings as were applied to the
actual data. These procedures were repeated using another set of 3000 simulated
point source maps based on the \citet{2011ApJ...742...24L} models, which predict a higher source
density with a mean of 3780 sources ($S>0.01$ mJy) per simulated image.

The positions of the extracted sources were matched with those of the 
simulated input sources for each of the 3000 simulations based on the two
different models. For a given extracted source, the matching input simulated 
source was chosen as the brightest source within a 5 pixel = $15''$ radius and 
with a brightness $>0.1$ mJy. In most cases there is only one source within 
$15''$, but infrequently a faint input source happens to lie closer to 
the extracted source position than a brighter input source that is the true
origin of the extracted source. 

Figure \ref{fig:S-vs-S} compares the brightnesses of the extracted sources with
the associated input source brightnesses. Results are binned in 0.1 mJy intervals,
and error bars indicate the standard deviation of the sources averaged in each bin.
For either set of simulations, we find that source boosting becomes significant at 
$S\lesssim1$ mJy. Consistent with the fluxes we measure for the extracted sources from the GDF, typical fluxes extracted from the simulations are greater than $\sim$400\,uJy/beam, as we have used the same detection criterion (at 2.99-$\sigma$) on our simulated data as on our observed map, and because our map has an average 1-$\sigma$ depth of around 135--140 mJy/beam. 
Figure \ref{fig:completeness} shows the completeness of the 
extracted sources, or the fraction of the input sources that are extracted. Whereas
essentially all sources with $S>1$ mJy are recovered by the source extraction,
the completeness drops to $\sim50\%$ at $S\sim0.4$ mJy for both models
of the source counts. Figure \ref{fig:position} shows the expected trend of
increasing positional errors with decreasing source brightness. The positions
of fainter sources are more strongly affected by noise and source confusion, consistent with Eq. 9. 
In Figure \ref{fig:position-eq-vs-sim} we compare the search radius given by Eq. 9 with the results from the simulated positional accuracy. The figure demonstrates that over the relevant range of fluxes, the calculated search radii are in remarkable agreement with simulations, thus justifying our reliance on them.
An assessment of the reliability of the source extractions is shown in Figure 
\ref{fig:reliability}, where we plot the fraction of sources at a given extracted
brightness that can be associated with corresponding input source in the simulations.
For either model, the reliability of the extracted sources begins to drop at $S<1.2$ mJy.
The drop is somewhat greater for simulations using the B\'ethermin et al. (2011) source
counts, which has an overall lower density of sources. As expected, the reliability drops more rapidly near the
detection threshold ($\sim$400 mJy/beam, given the average 1-$\sigma$ depth around 135--140 mJy/beam over the full map). Note, that because the reliability depends strongly on the shape of the unknown actual number counts, especially near the chosen detection threshold, the confidence levels shown in Table~\ref{tab:source_fluxes}, and calculated using Eq.~\ref{eq:confidence}, serve as our guiding estimates of the source reliabilities in the absence of prior knowledge of the actual source counts.
Figure \ref{fig:false} shows the
same information in a slightly different way. Histograms of the extracted source brightnesses
are compared to the histograms of the input simulation source brightnesses (when they exist).
The ratio of the two histograms yields the reliability that was plotted in 
Figure \ref{fig:reliability}. The simulations lead us to expect an average of 
1.34 false detections in our field if the source counts are similar to the 
\citet{2011A&A...529A...4B} model, or 0.55 sources per field if the sources are more numerous 
as in the \citet{2011ApJ...742...24L} model.
Finally Figures \ref{fig:simhistsa}  and  \ref{fig:simhistsb} show histograms of the total number of sources extracted
and histograms of the number of false sources extracted across the 3000 simulations. The histograms
tend to show a slight positive tail with respect to Poisson distributions with the 
same mean number of sources extracted.

\section{Discussion}
\subsection{2~mm number counts}
\label{sec:counts}
\begin{figure} 
   \centering
   \includegraphics[width=3.3in]{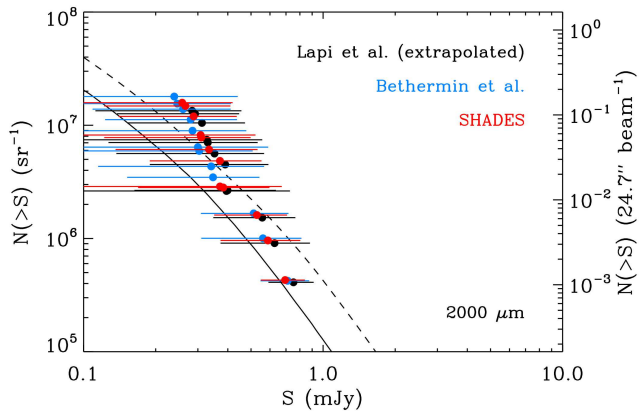} 
   \caption{The number counts 
   \hbox{$N(>S)$} as a function of the deboosted 2mm fluxes ($S$). 
   The black symbols show the deboosting using the \citet{2011ApJ...742...24L} model 
   extrapolated down to 0.01 mJy. The blue and red show
   the deboosted number counts when the \citet{2011A&A...529A...4B} and the 
   SHADES \citep{2006MNRAS.372.1621C} models are used to calculate the deboosting.
   The solid line is an interpolation of the 1.38 and 2.1 mm model number 
   counts of \citet{2011A&A...529A...4B}, and the dashed line is the model from 
   \citet{2011ApJ...742...24L}. The right-hand axis labels the counts in terms of number of sources per beam
   for convenience in assessing confusion. The counts were corrected for the signal-to-noise dependent effective area for source extraction in the map, as well as for the expected number of false detections from Table \ref{tab:source_fluxes}.
   \label{fig:NgtS}}
\end{figure}

Figure \ref{fig:NgtS} depicts the observed cumulative number counts, \hbox{$N(>S)$}, as a function of the 
deboosted 2 mm flux density, $S$, compared with the predicted galaxy 
counts in \citet{2011A&A...529A...4B} (the 
solid line is a power-law interpolation between their 2.1 and 1.3 mm models), and the model 
from \citet{2011ApJ...742...24L} (dashed line). These counts are not binned due to the small number of sources.
Instead, as in e.g. Borys et al. (2003), we plot the number of sources 
at each deboosted flux density, divided
by the effective area for the detection of sources of a given flux density. 
Following \citet{2006MNRAS.372.1621C}, the effective area is calculated
as the product of the maximum area of field and the fit to the completeness function 
(Fig. \ref{fig:completeness}), although in our case the functional form is simpler and 
involves only one free parameter. The effective area ranges from 0.21 to 0.90 times the maximum area.

Instead of plotting the cumulative number counts as a simple stair-stepped line, we include the 
uncertainties for each of the deboosted flux densities. 
Figure \ref{fig:NgtS} also demonstrates  that the cumulative number counts are independent of the three deboosting models were used, considering the uncertainties
in the deboosted flux densities. We note that the counts are somewhat steeper than the models.


On a cautionary note, our number counts represent 
measurements from a small patch of the sky, therefore allowing for quite some uncertainty 
in terms of cosmic variance, which in the context of AzTEC 1.1~mm number counts is being discussed in 
\citet{2012MNRAS.423..575S}. 

\subsection{Source properties and associations}

The GISMO 2-mm datum typically adds an important long-wavelength constraint to the FIR Spectral-Energy Distribution (SED) of a distant ($z>$2) infrared galaxy, provided it is measured with sufficient precision. When it is the only datum on the Rayleigh-Jeans side of the thermal greybody spectrum, it provides a critical constraint on the infrared luminosity of the galaxy
and the cold star-forming dust mass. And, in combination with other Rayleigh-Jeans data (such as 850\,$\mu$m or 1.2\,mm), the 2-mm point can furthermore constrain the temperature and the dust emissivity index $\beta$, which is a diagnostic of the physical geometry of the dust grains (see e.g.\ Yang et al.\ 2007).

Because the large statistical uncertainties associated with the underlying source fluxes in case of the low signal-to-noise (SNR$<$5) detections, where statistical flux boosting is significant (section \ref{sec:deboost}), we cannot provide accurate 2-mm photometry individually for most of the detected sources, except for GDF-2000.1 and GDF-2000.3. However, we can use the GISMO data collectively to determine useful constraints for the population as a whole.

To fit the radio-to-FIR SEDs of our sources, we relied on the analytic temperature-distribution models of \citet{2010ApJ...717...29K}, which assumes a power law distribution ($dM_d / dT \propto T^{-\gamma}$) of dust components above a dominant cold-temperature component at $T_c$. For the fitting we have assumed a characteristic emission diameter of 2\,kpc, typical to SMGs, and a dust emissivity index of $\beta$=1.5, typical for starbursts (Kov\'{a}cs et al.\ 2006, 2010). We use all available continuum data at wavelengths longer than 15\,$\mu$m (rest frame $3\,\mu$m), a regime dominated by thermal dust (millimeter wavelengths to FIR) and synchrotron radiation (at the radio wavelengths). We assume a 10\% calibration uncertainty for all measured bands, added in quadrature to the reported detection uncertainties.

The collective fit to all GISMO sources with sufficient photometry (sources 1, 3, 6, 8, 9, 11, 13 and 14) yields $\gamma$=7.24$\pm$0.23, in excellent agreement with local starburst galaxies (see Kov\'{a}cs et al.\ 2010). Therefore, for the two most significant individual sources, which we discuss in the following, we fix $\gamma$=7.2, and fit only the cold-component temperature, dust mass and the radio-FIR correlation constant $q_L$ (Kov\'{a}cs et al. 2010) or $q_{IR}$ (Ivison et al.\ 2010). For the synchrotron spectral index ($S_\nu \propto \nu^{-\alpha}$) we assume $\alpha$=0.75.

\subsubsection{GDF-2000.1 \& GDF-2000.3}

GDF-2000.1 and GDF-2000.3 are the two most significant GDF detections with counterparts. Our SED fits for these sources follow the method described in \citet{2010ApJ...717...29K}. The plots show IRAC data (at $\lambda < 8~\mu$m), but those were not used for the fit. A summary of the fitted parameters is given in Table 6. 

\begin{figure}
\includegraphics[angle=0,width=3.3in]{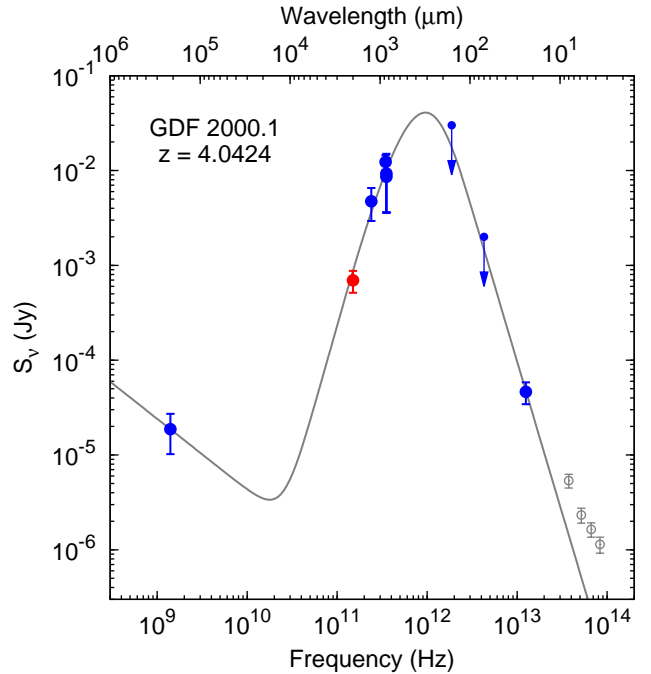}
\caption{ SED of GDF-2000.1. The red point at 2mm is the GISMO measurement. The solid line is the model fit to the SED (see text and Table 6), but does not include the 3.6 - 8 $\micron$ IRAC data as constraints.} 
\label{GDF-SED1}
\end{figure}

The detection of GDF-2000.1, has a signal to noise ratio $\sigma > 5$ and a statistical detection confidence level of 100\%. It is associated with 
AzGN03 (also dubbed: GN1200.02, GN850.10, MM J123633+621407, SMM 
J123633.8+621408), and GN10 in \citet{2005MNRAS.358..149P} and \citet{2006MNRAS.370.1185P}. This is the source 
discussed in, e.g., \citet{2008ApJ...673L.127D}, and in \citet{2009ApJ...695L.176D} who determine its redshift as $z=4.04$ through measurements of CO(4-3), the redshift we assume for this source. We note that 
\citet{2004AJ....127.3121W} report a possible counterpart at $z=1.34476$, flagged as ``very secure $z$'', with $>99$\% confidence, centered $\sim3''$ north of  the nominal center of GDF2000.1 coordinates. Fig. \ref{GDF-SED1} shows the SED of GDF-2000.1, using all available flux information of this source at other wavelengths.

\begin{figure}
\includegraphics[angle=0,width=3.3in]{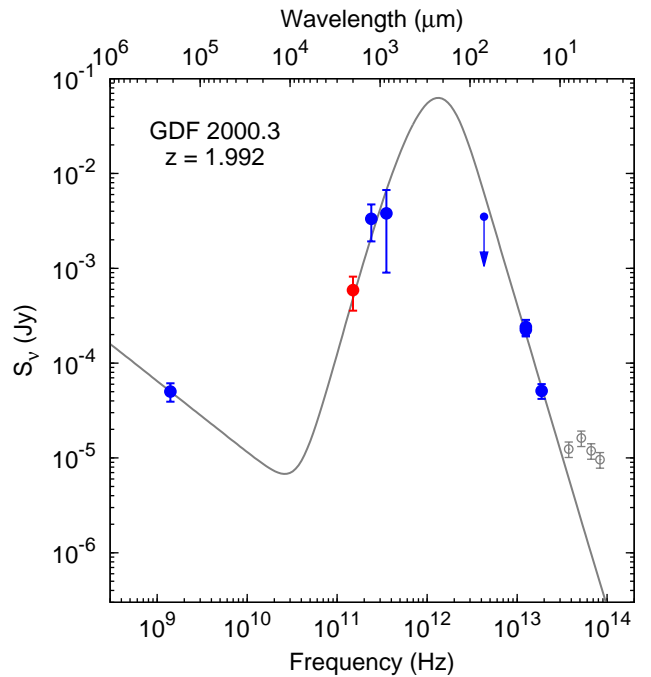}
\caption{ SED of GDF-2000.3. Symbols and lines as in Fig. \ref{GDF-SED1}.} 
\label{GDF-SED3}
\end{figure}

GDF-2000.3, is detected with a signal to noise ratio of  $\sigma > 4$. With 99\% the confidence level for detection is very high. GDF-2000.3 has the SCUBA counterpart GN850.39 and MAMBO/AzTEC counterpart AzGN07. The (sub)millimeter counterparts are  described in \citet{2006MNRAS.370.1185P} and \citet{2008MNRAS.389.1489G}. Two sources with known redshifts are positionally consistent with our measured position: GOODS J123711.98+621325.7 with $z$ = 1.992, and SMM J123711.9+621331, with $z$ = 1.990. Since there are no  other IR sources similarly close we consider those plausible counterparts and use the redshift for our SED fit (Fig. \ref{GDF-SED3}).

GDF-2000.1 and GDF-2000.3 are HLIRGs, with $L>10^{13} L_{sun}$, and $SFR > 1000 M_{sun}$/yr. The estimated optical depths ($\sim 1$) at the peak of the emission are typical to SMGs. Both q values are significantly higher than the radio-FIR correlation for SMGs (\citet{2006ApJ...650..592K} and \citet{2010ApJ...717...29K} both found $<q_L> \sim 2.13$ for SMGs, with a scatter of 0.12 dex only). Thus, the excess far-infrared emission might need to be explained by the presence of an additional significant heating source besides stars, possibly a powerful AGN.  This is unlike the bulk of the known SMG population, where AGNs, although often present, are not significant contributors ($<$20\%) to the infrared emission. 

\subsubsection{GDF-2000.6}

GDF2000.6 
is HDF850.1 ($z=5.2$), the most prominent among the first ever identified SCUBA Deep Field sources \citep{1998Natur.394..241H}. The observed GISMO position and the deboosted 2~mm flux are consistent with the data published in \citet{2012Natur.486..233W}. 

\subsubsection{Other GDF sources with counterparts}

We identify AzGN07 as the counterpart for GDF-2000.8,  AzGN08 as the counterpart for GDF-2000.11, and GN850.21 and GN1200.29 as counterparts for the low S/N source GDF-2000.13.
We associate CXO J123627.53+621218.0  (with a photometric redshift $z$ = 4.69) and AzGN10 as  counterparts   of GDF-2000.14, since the positions of these two sources are essentially identical, and in
 very good agreement with the GISMO detection. Furthermore, the high photometric redshift value for the CXO source plus the observed 1~mm flux makes a 2~mm detection at our sensitivity level very likely. Finally, we identify  GN850.28 as counterpart to GDF-2000.15.

\section{Summary and Conclusion}

We have obtained a $7'$ diameter 2 mm continuum deep field map centered on the HDF. The rms in the inner part of the  map is $\sim135 \mu$Jy/beam.
The noise in the un-smoothed data very closely follows a Gaussian distribution, indicating its random nature and validating probabilistic source extraction statistics.

We detect 12 sources plus 3 false, negative ``detections''.  A statistical analysis of the data  predicts 2 false detections. 

5 of the detected 12 sources have known (sub)millimeter counterparts, including HDF850.1, the first submillimeter galaxy detected by SCUBA. Three more low signal to noise sources have been identified through cross correlation with existing (sub)mm data. The mean redshift of all 7 of the counterparts with known redshifts is $\bar{z} = 3.3$, the median redshift  of those sources, which at this low number of sources is probably a better estimate,  is $\tilde{z} = 2.91 \pm 0.94$.

Of the remaining 7 detected sources which have no (sub)millimeter counterpart, statistically we expect 5 to be real. 

The jackknife test of the smoothed data shows an almost perfect Gaussian distribution for the S/N histogram. 
The S/N histogram of the normally processed, smoothed data shows a clear excess beyond a Gaussian distribution, which mostly can be contributed to 12 astronomical sources in the field. 
A small symmetric excess remains after the resolved sources are subtracted from the image. This most likely indicates the presence of confusion noise in our data. 


\acknowledgements

We would like to thank Carsten Kramer, Santiago Navarro, David John, Albrecht Sievers, and the entire IRAM Granada staff for their support during the instrument installation and observations.
IRAM is supported by INSU/CNRS (France), MPG (Germany) and IGN (Spain). This work was supported through NSF ATI grants 1020981 and 1106284. 

{\it Facility:} \facility{IRAM:30m (GISMO)}



\appendix

\begin{figure} [t]
   \centering
      \includegraphics[width=1.7in]{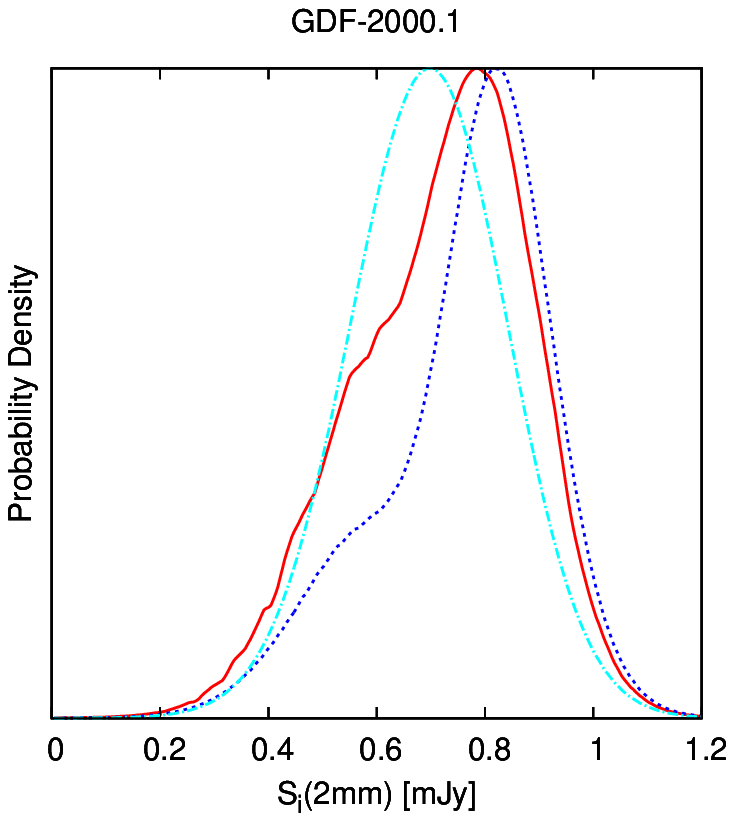} 
      \includegraphics[width=1.7in]{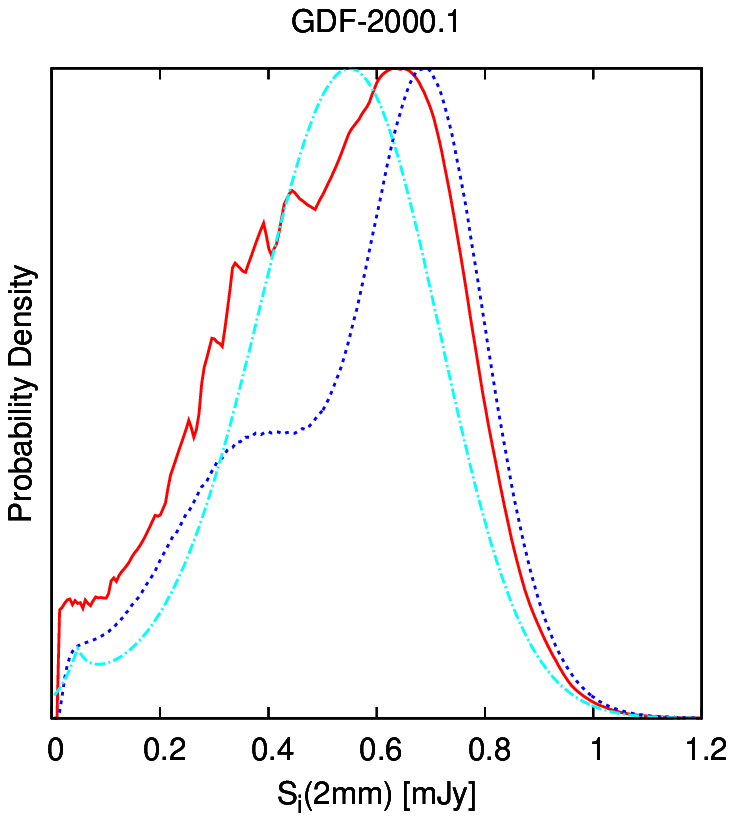}    
      \includegraphics[width=1.7in]{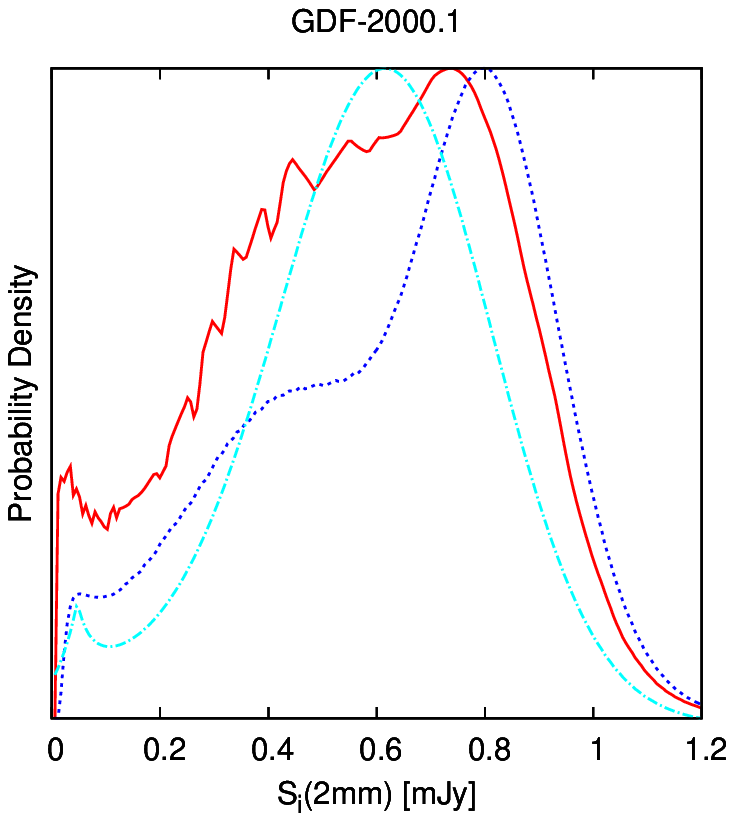} 
      \includegraphics[width=1.7in]{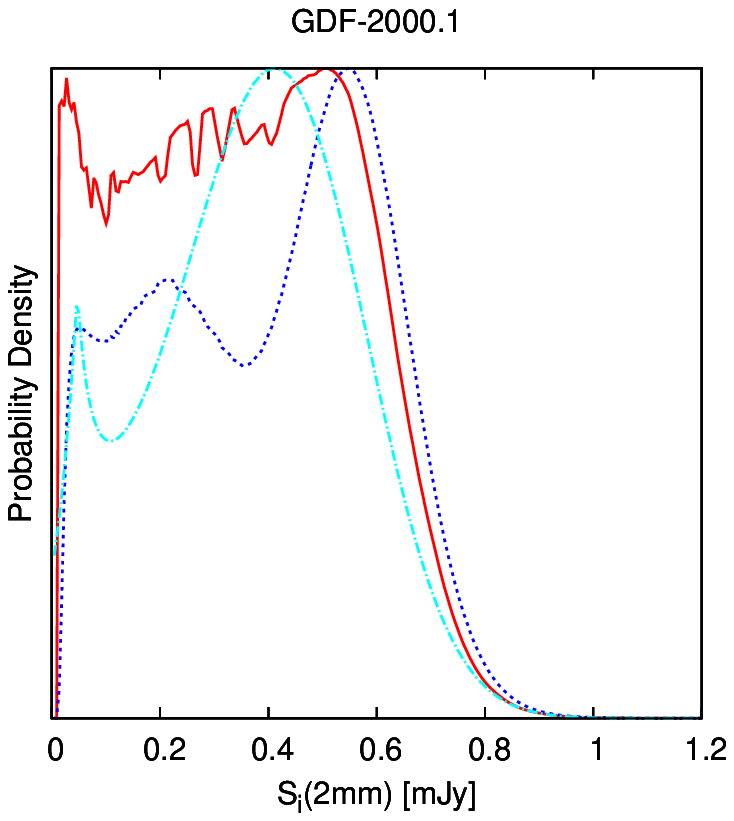} 
      \includegraphics[width=1.7in]{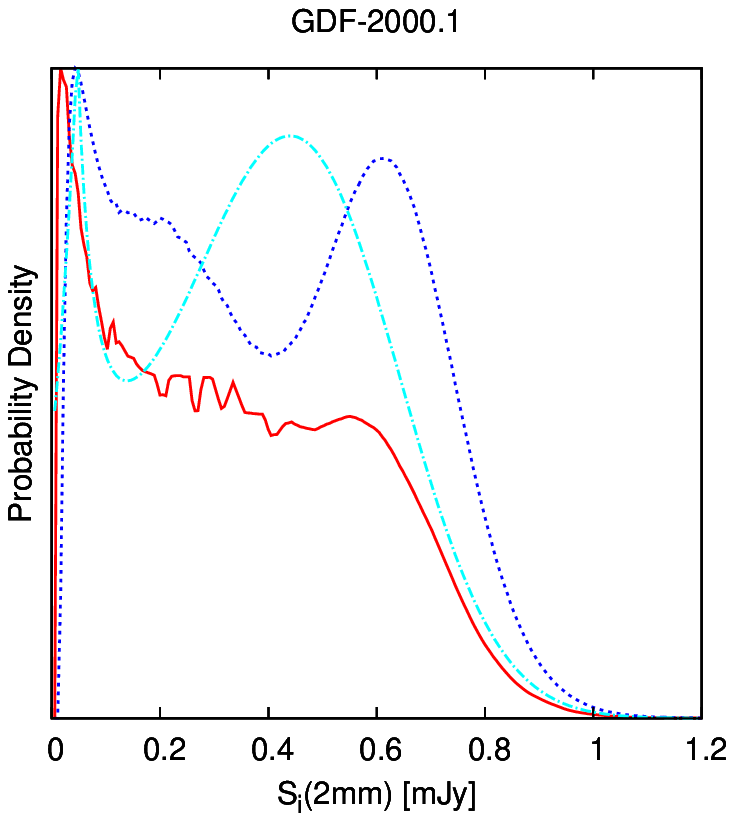} 
      \includegraphics[width=1.7in]{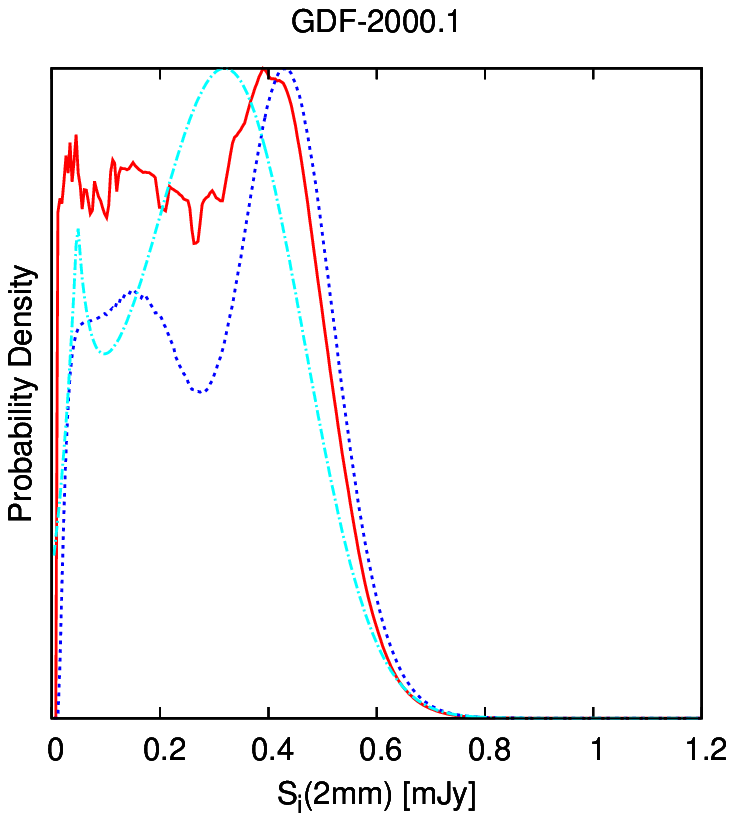} 
      \includegraphics[width=1.7in]{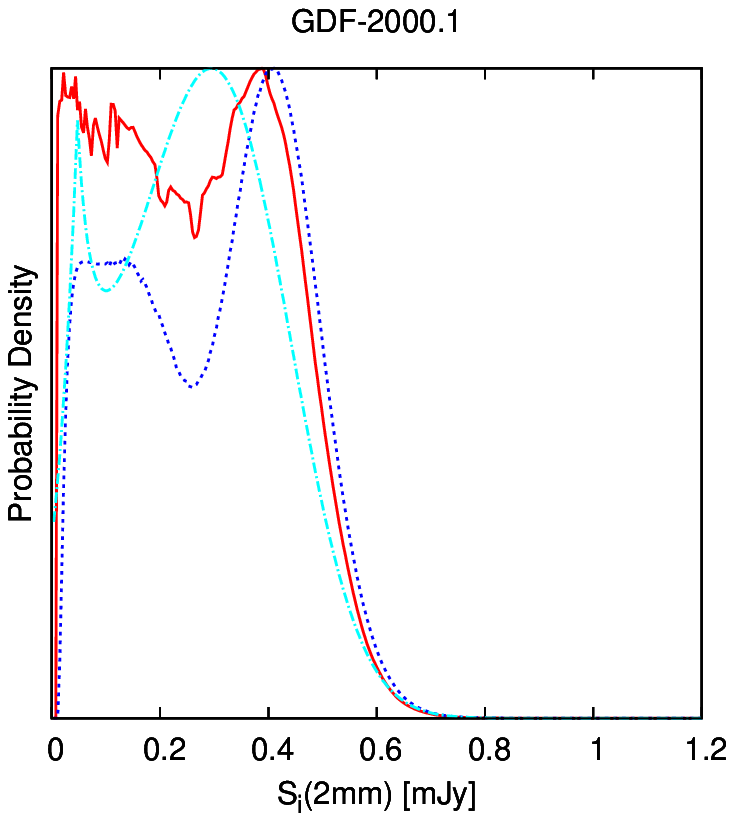} 
      \includegraphics[width=1.7in]{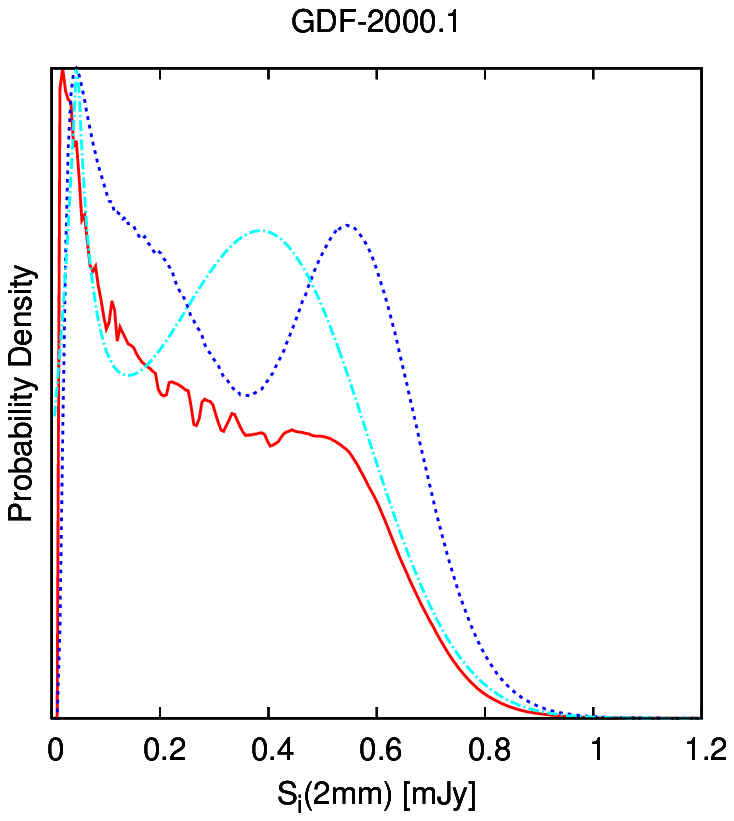} 
      \includegraphics[width=1.7in]{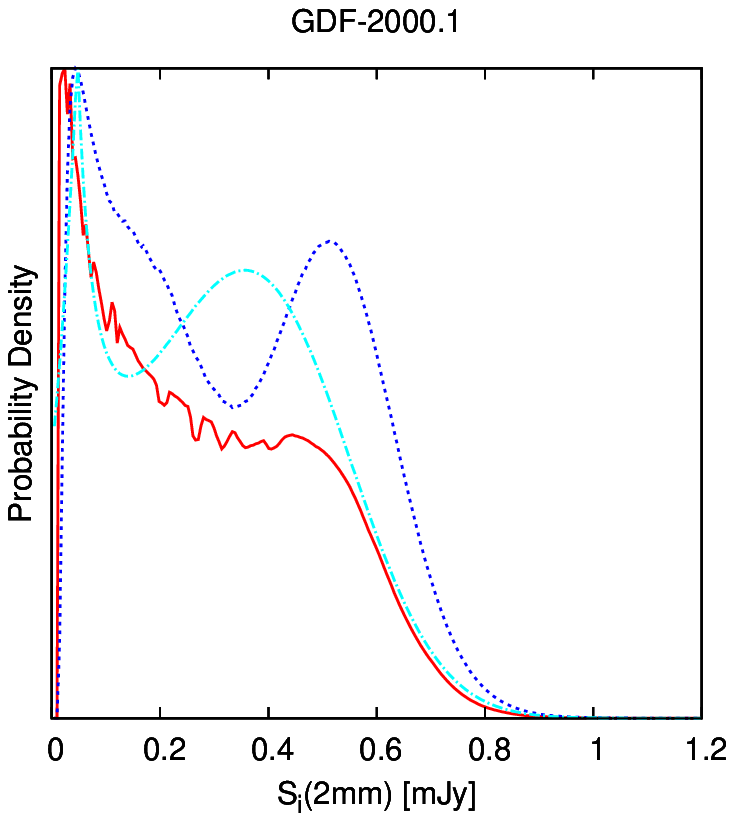} 
      \includegraphics[width=1.7in]{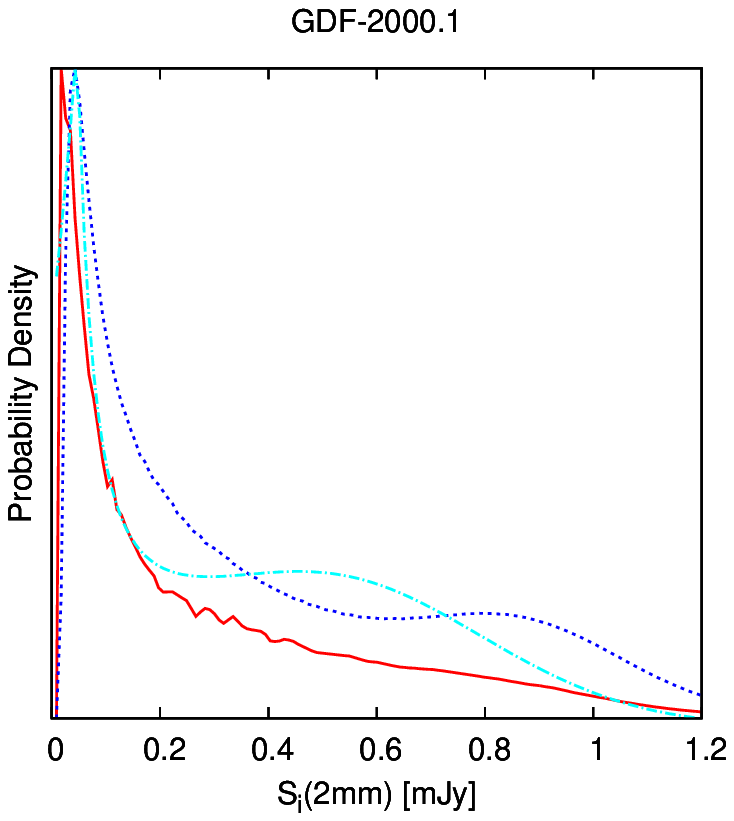} 
      \includegraphics[width=1.7in]{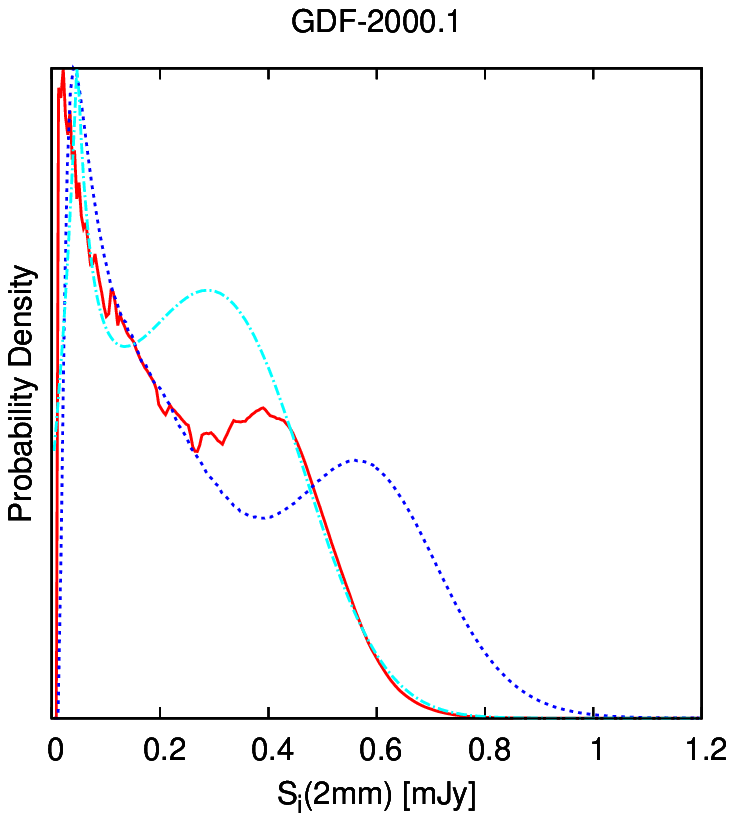} 
      \includegraphics[width=1.7in]{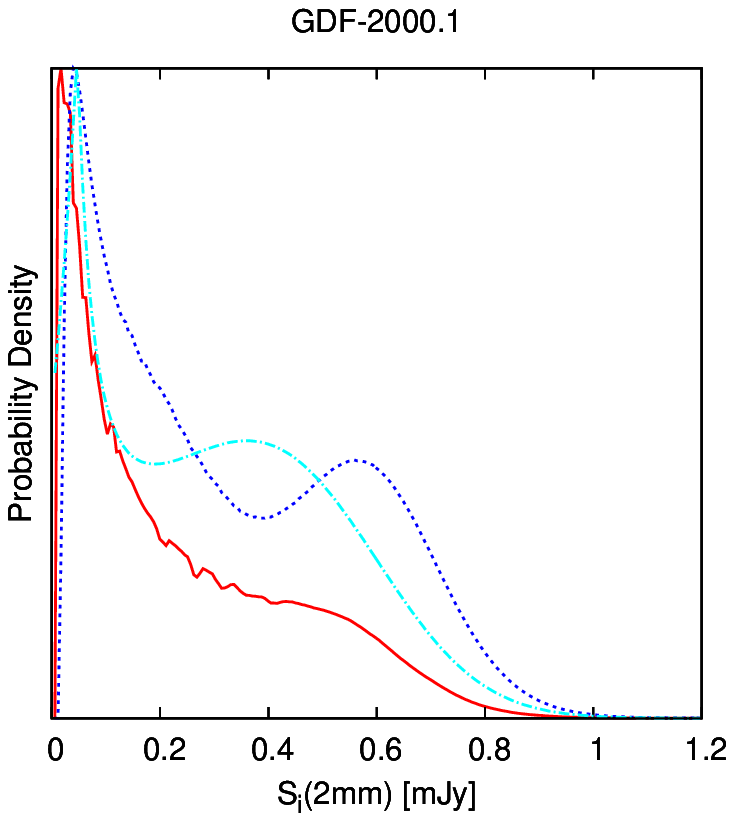} 
      \includegraphics[width=1.7in]{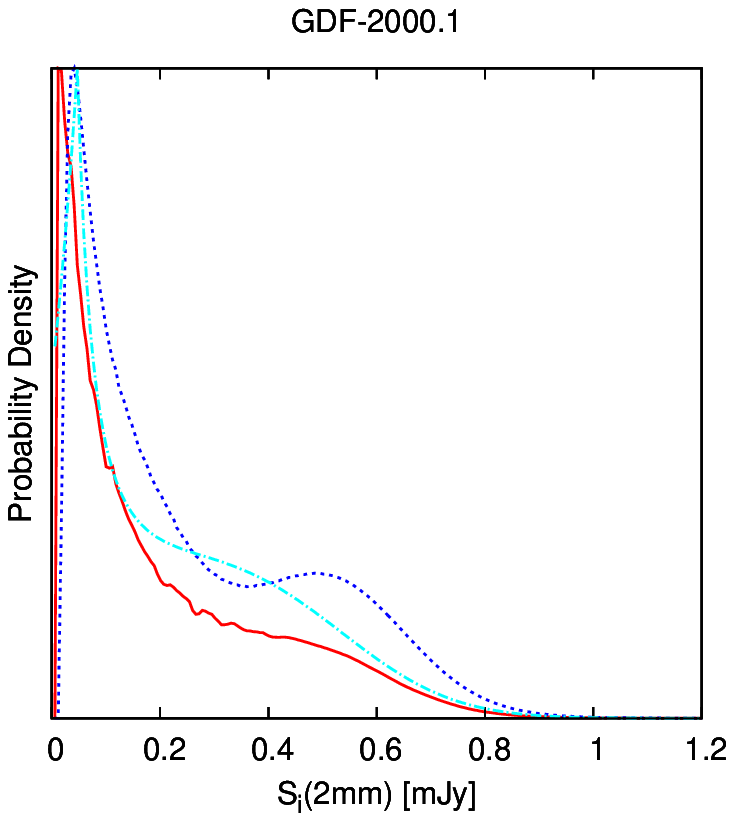} 
      \includegraphics[width=1.7in]{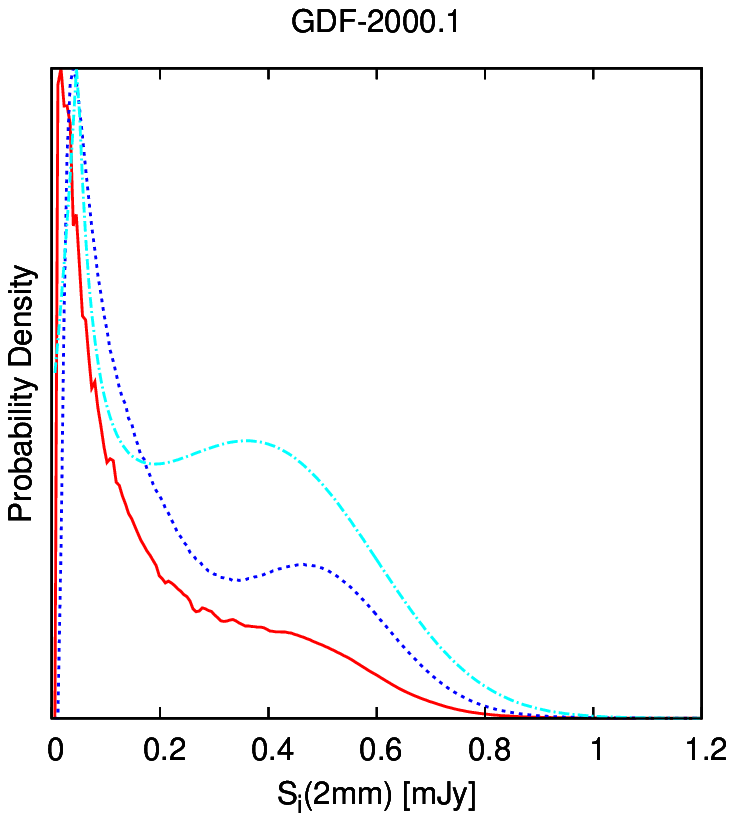} 
      \includegraphics[width=1.7in]{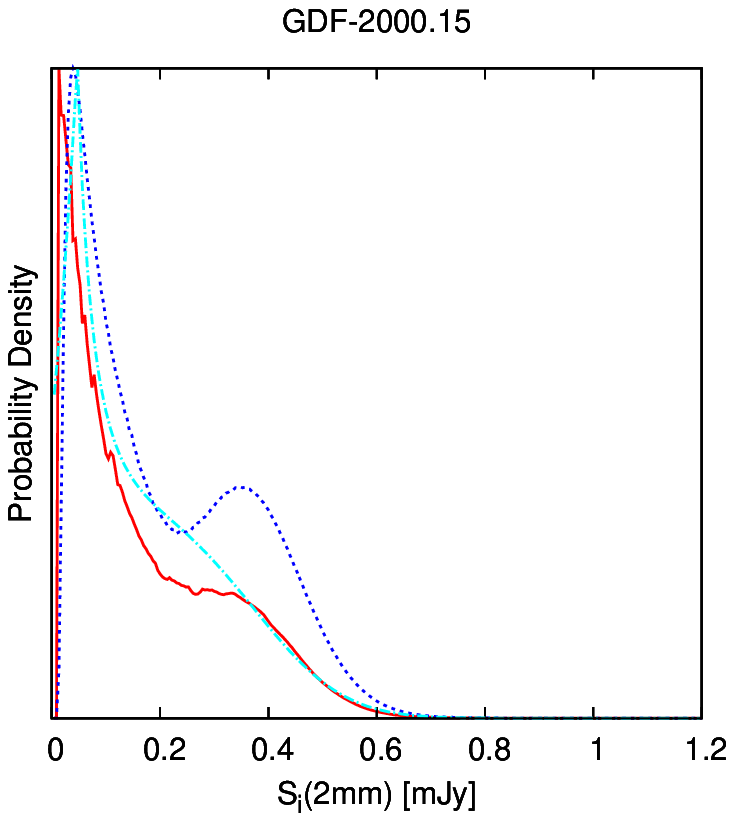}  \\
   \caption{Posterior probability distributions for all GISMO sources,
indicating the Bayesian probability densities that the detected
source flux arises from an intrinsic source flux of $S_i$. The
probabilities account for up to 3 overlapping resolved sources
contributing to the observed flux, and take confusion at the faint
end into account. The distributions are shown for the number count
models of \citet{2011A&A...529A...4B} in red/solid, the extended \citet{2011ApJ...742...24L} counts in blue/dashed, and the scaled broken powerlaw
SHADES counts in cyan/dashed-dotted.
Note, that the jagged curves resulting from the \citet{2011A&A...529A...4B} counts are not a property of our algorithm, but are inherent to the input counts model.}
   \label{fig:S_vs_S}
\end{figure}

Figure \ref{fig:S_vs_S} shows the 
posterior probability distributions for all GISMO sources,
indicating the Bayesian probability densities that the detected
source flux arises from an intrinsic source flux of $S_i$. The
probabilities account for up to 3 overlapping resolved sources
contributing to the observed flux, and take confusion at the faint
end into account.  Since our deboosting method is based on resolved sources only, no additional zero-level adjustment is necessary. 
As a result, the distribution naturally does not extend to negative fluxes, as is evident in the figure. The distributions are shown for the number count
models of \citet{2011A&A...529A...4B} in red/solid, the extended \citet{2011ApJ...742...24L} counts in blue/dashed, and the scaled broken powerlaw
SHADES counts in cyan/dashed-dotted.

\end{document}